\documentclass[aps,prd,superscriptaddress,floatfix,nofootinbib,preprintnumbers]{revtex4}
\usepackage{amsmath,multirow}
\usepackage{graphicx,tabularx,slashed}
\usepackage{url}
\usepackage{footmisc}
\usepackage{hyperref}

\newcommand{\mg}{{\sc MadGolem \,}}

\newcommand\one{\leavevmode\hbox{\small1\normalsize\kern-.33em1}}

\newcommand{\lag}{\mathcal{L}}

\newcommand{\qqquad}{\qquad \qquad}

\newcommand{\msbar}{\ensuremath{\overline{\text{MS}}}}

\newcommand{\go}{\tilde{g}}

\newcommand{\gev}{{\ensuremath\rm GeV}}

\def\slashchar#1{\setbox0=\hbox{$#1$}           
   \dimen0=\wd0                                 
   \setbox1=\hbox{/} \dimen1=\wd1               
   \ifdim\dimen0>\dimen1                        
      \rlap{\hbox to \dimen0{\hfil/\hfil}}      
      #1                                        
   \else                                        
      \rlap{\hbox to \dimen1{\hfil$#1$\hfil}}   
      /                                         
   \fi}

\def\eg{{\sl e.g.} \,}
\def\ie{{\sl i.e.} \,}

\newcommand{\myrbox}[1]{\parbox{2.5cm}{\resizebox{2.5cm}{!}{#1}}}

\newcommand{\lo}{\ensuremath{l_8}}
\newcommand{\lobar}{\ensuremath{\bar{l}_8}}
\newcommand{\eo}{\ensuremath{e^+_8}}
\newcommand{\eobar}{\ensuremath{e^-_8}}
\newcommand{\nuo}{\ensuremath{\nu_8}}
\newcommand{\nuobar}{\ensuremath{\nu_8}}
\newcommand{\rself}{\ensuremath{\hat{\Sigma}}}

\begin{document}


\title{Looking for leptogluons}

\author{Dorival Gon\c{c}alves-Netto}
\affiliation{Institut f\"ur Theoretische Physik, Universit\"at Heidelberg, Germany}
\affiliation{Max-Planck Institut f\"ur Physik, Munich, Germany}

\author{David L\'opez-Val}
\affiliation{Institut f\"ur Theoretische Physik, Universit\"at Heidelberg, Germany}

\author{Kentarou Mawatari}
\affiliation{Theoretische Natuurkunde and IIHE/ELEM, Vrije Universiteit Brussel, Belgium \\ 
and International Solvay Institutes, Brussels, Belgium}

\author{Ioan Wigmore}
\affiliation{SUPA, School of Physics \& Astronomy, The University of Edinburgh, Scotland, UK}

\author{Tilman Plehn}
\affiliation{Institut f\"ur Theoretische Physik, Universit\"at Heidelberg, Germany}

\begin{abstract}

We present search results based on next-to-leading order predictions for the
pair production of color-adjoint leptons at the LHC.  Quantum effects
are sizable, dominated by pure QCD corrections, and sensitive to
threshold effects.  We illustrate the stabilization of scale
dependences and confirm an excellent agreement between fixed-order and
multi-jet predictions for representative distributions.  Finally, we
examine the trademark collider signatures of leptogluon pairs. Based
on the CMS leptoquark search we derive a mass bound of 1.2-1.3~TeV for
charged leptogluons, significantly improving the constraints available
in the literature.
\end{abstract}

\maketitle

\tableofcontents 

\newpage

\section{Introduction}
\label{sec:intro}

Leptogluons are color-adjoint fermions with non-zero lepton number. 
We can categorize them as strongly interacting partners of the Standard Model leptons.
Leptogluons may arise, for instance, as a manifestation of compositeness~\cite{compositeness}.
Proposals along these lines are motivated by the flavor and high-scale gauge structure of the Standard Model which
suggests a relation between quarks and 
leptons.
This can be explained, for instance, if both types of particles share
a common substructure and emerge as bound states of fundamental constituents.
%
These building blocks will be bound together 
by a new confining force. The onset of a confinement
scale $\Lambda$ will characterize the typical masses of the new heavy composites. 
If the basic quark and lepton constituents turn out to be charged
under the Standard Model gauge group the new composite quarks and leptons will inherit
these gauge charges and 
will thus be pair-produced through their ordinary couplings to gauge bosons.
Alternatively, they can 
be produced in association with their Standard Model counterparts, 
via higher dimensional operators which govern 
the radiative transitions between both.

\smallskip{} 
Historically, the seminal ideas on compositeness~\cite{seminal} cristallized  
into a paradigmatic class of models, the so-called preon models~\cite{preons}. 
With the term \emph{preon}, as coined by Pati and Salam, 
one denotes these more fundamental building
blocks of Nature, whose condensates make up the ``elementary'' quarks and leptons,
together with a tower of excited states. 
For example,
in the framework of the so-called fermion-scalar models, the Standard Model leptons
emerge as bound states from the combination of a fermionic preon and a scalar
anti-preon $l \equiv (F\bar{S})$; if these constituents
are charged under $SU(3)_c$, this naturally leads to
leptonic composites of the sort $3\otimes\bar{3} = 1\oplus8$:
the color-singlet state can then be identified with an ordinary lepton,
while the color-octet lepton corresponds
to a leptogluon, \eg the color-octet electron $e^\pm_8$. 
This argument illustrates how novel fermionic states transforming under higher
$SU(3)_c$ representations 
serve as telltale predictions of compositeness models. 
Scalar or vector leptoquarks are another example of such heavy colored resonances~\cite{Buchmuller:1986zs}.
More generally,~TeV-scale strongly interacting sectors are typically linked to 
some kind of compositeness~\cite{Luty:1996jd,Agashe:2004rs,Gabella:2007cp,Buckley:2012ky}. Their 
collider imprints, for instance the production of exotic objects such as e.g. excited leptons \cite{excited-leptons},
quark sextets and decouplets~\cite{signatures2}, 
have been gathering attention in the recent years. 

\bigskip{} 
Phenomenological studies of leptogluon production are available 
for a diversity of frameworks, ranging from dedicated calculations
of leptogluon 
pair production at $e^+ e^-$, $ep$, and 
$pp(\bar{p})$ colliders~\cite{phenostudies}
to generic approaches where
the characteristic signatures of exotic colored states are portrayed, not only 
at colliders~\cite{signatures,signatures2} but also for precision physics~\cite{GonzalezGarcia:1996rx}. 
Very recently, studies on the prospects of leptogluon searches
at the LHC ~\cite{Mandal:2012rx} and the LHeC \cite{Sahin:2013vha} have been published.

\smallskip{}
Experimentally, the
current $3\sigma$, model-independent mass bounds from collider searches 
yield $m_{e_8} > 86$~GeV for a stable
charged lepton octet; and $m_{\nu_8} > 110$~GeV, in the case
of its neutral analoge $\nu_8$~\cite{pdg}. 
Dedicated analyses by JADE and H1 
more than two decades ago have extracted leptogluon mass
bounds attached to specific compositeness scale choices,
typically in the range of $m_{e^{\pm}_8} \sim 100-200$~GeV
for $\Lambda \sim 1-2$~TeV~\cite{phenoexp1}.

\bigskip{}
In this paper we analyse the production of leptogluon pairs to next-to-leading order (NLO) QCD.
We compute the total leptogluon pair production rates
and examine in detail 
the different quantum effects.
The reduced theoretical uncertainties we illustrate via the
stabilization of the NLO predictions with respect to
variations of the (unphysical) renormalization and 
factorization scales.
We compute representative kinematic distributions to NLO and compare them to 
independent results based on multi-jet merging. 
All these NLO calculations we carry out with the help of the fully
automated package {\sc MadGolem}~\cite{madgolem,sqn1,sgluon,susypair}.
Finally, we explore 
the characteristic collider signatures of leptogluon pair production 
and discuss the prospects for leptogluon searches at the LHC.


\section{Leptogluon pairs to NLO}
\label{sec:nlo}

We analyse the phenomenology of leptogluons at the LHC assuming a minimal extension of 
the Standard Model. We entertain the existence of an additional generation of lepton octets
which consists of one electrically charged, color-adjoint (Dirac) lepton ---
the \emph{electron/positron octet} $e^{\pm}_8$ --- and its neutral
companion, a \emph{neutrino octet} $\nu_8$. 
Generically,
we shall label these fields as $l_8 \equiv e^{\pm}_8, \nu_8$.
For our analysis,
no assumption needs to be made on their possible weak charges. 
For concreteness, and unless otherwise stated, we shall assume both states to be
mass degenerate. This setup provides a 
minimal, but representative framework in which to explore the trademark collider
imprints of leptogluons. Interestingly, 
neutrino octets from compositeness scenarios
and gluino fields from supersymmetric models share the same quantum numbers (barring 
the lepton number). In this sense, neutrino octets will portray very similar aspects
as gluinos, with no connection to the 
supersymmetric nature of the latter.  

\smallskip{}
Leptogluon interactions at the LHC are entirely driven by QCD. 
As $SU(3)_c$ adjoints leptogluons couple to gluons 
through the covariant derivative 
$(D_\mu)_{AB} = (\partial_\mu)_{AB} + i\,g_s\,(T^C\,A^C_\mu)_{AB}$, 
where $A_\mu^C$ denotes the gluon field, $g_s$
the strong coupling constant, and $T^C$ the adjoint $SU(3)_c$ generators. We write
these generators in terms of the structure constants 
$T^A_{BC} = -if^{ABC}$.
The Lagrangian that describes the $g l_8 l_8$ coupling then yields
\begin{equation}
 \lag \supset \overline{\psi}_{l_8}\,(i\slashed{D} - m_{l_8})\,\psi_{l_8} \supset -g_s\,f^{ABC}\,\overline{\psi}^A_{l_8}\,\gamma^\mu\,\psi^B_{l_8}\,\,A^{C}_\mu \; .
\label{eq:lagtree}
\end{equation}

\begin{figure}[t!]
\begin{center}
\includegraphics[width=0.13\textwidth]{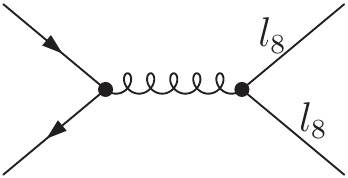} 
\hspace*{0.1\textwidth} 
\includegraphics[width=0.37\textwidth]{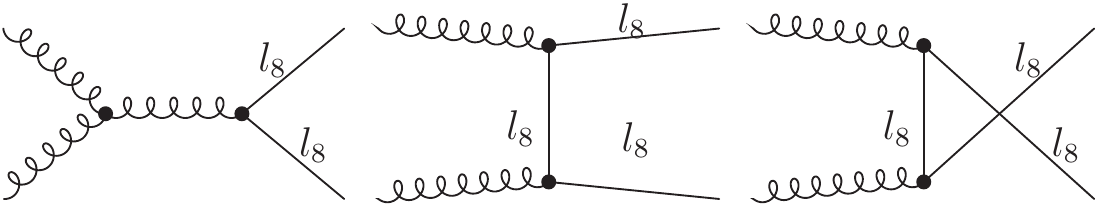}
\end{center}
\caption{Feynman diagrams for the pair production of leptogluons
 to leading order. We represent both partonic subprocesses:
 quark-antiquark annihilation $q\bar{q} \to \lo\lobar$ and gluon fusion $gg \to \lo\lobar$.}
\label{fig:born}
\end{figure}

\bigskip{} 
There are two partonic subprocesses which contribute to pair production of leptogluons at the LHC:
quark-antiquark annihilation and gluon fusion (cf. Figure~\ref{fig:born}):  
\begin{alignat}{5}
pp(q\bar{q}) \to \lo\lobar \qquad \text{and} \quad  pp(gg) \to \lo\lobar \; ,
 \label{eq:lochannels}
\end{alignat}
where $\lo\lobar$ can either stand for the production of a charged $\eobar\eo$
or a neutral $\nuo\nuobar$ leptogluon pair. 
Throughout our analysis we set the 
central renormalization and factorization scales to the average
final state mass $\mu^0 \equiv \mu_{R,F} = m_{\lo}$, which leads to stable perturbative
results~\cite{prospino_sqgl,prospino_stop}. Unless stated otherwise, we fix the LHC energy to $\sqrt{S} = 8$~TeV, while the default leptogluon mass choice is 
$m_{\lo} = 1$~TeV. As for the parton densities we use 
CTEQ6L1 and CTEQ6M~\cite{cteq} with five active flavors. 
The corresponding values for the strong coupling
constant $\alpha_s$ are computed 
with consistent values of $\alpha_s(\mu_R)$ obtained via two-loop
running from $\Lambda_\text{QCD}$ to the required renormalization scale $\mu_R$, again in the five active flavor scheme.

\subsection{Production rates}
\label{ssec:rates}

\begin{table}[b]
\begin{center}
 \begin{tabular}{c|ccc|ccc|ccc} \hline 
$pp \to e^+_8\,e^-_8$ & \multicolumn{3}{c|}{$\sqrt{S} = 7$~TeV}  & \multicolumn{3}{c|}{$\sqrt{S} = 8$~TeV} &  \multicolumn{3}{c}{$\sqrt{S} = 14$~TeV} \\ \hline
$m_{l_8}\,$ [GeV] & $\sigma^\text{LO}$ [pb] & $\sigma^\text{NLO}$ [pb] & $K$ & $\sigma^\text{LO}$ [pb] & $\sigma^\text{NLO}$ [pb] & $K$
& $\sigma^\text{LO}$ [pb] & $\sigma^\text{NLO}$ [pb] & $K$\\ \hline
500 & $2.39 \times 10^{0}$ & $4.77 \times 10^{0}$ & 2.00 & $4.35 \times 10^{0}$ & $8.61 \times 10^{0}$ & 1.98 & $3.84 \times 10^{1}$ & $6.84 \times 10^{1}$ & 1.78 \\ 
700 & $1.86 \times 10^{-1}$ & $3.84 \times 10^{-1}$ & 2.06 & $3.85 \times 10^{-1}$ & $7.89 \times 10^{-1}$ & 2.05 & $4.98 \times 10^{0}$ & $9.08 \times 10^{0}$ & 1.83 \\ 
900 &  $2.13 \times 10^{-2}$ & $4.42 \times 10^{-2}$ & 2.08 & $5.02 \times 10^{-2}$ & $1.04 \times 10^{-1}$ & 2.08 & $9.20 \times 10^{-1}$ & $1.76 \times 10^{0}$ & 1.91 \\ 
1100 & $2.88 \times 10^{-3}$ & $6.03 \times 10^{-3}$ & 2.09 & $8.04 \times 10^{-3}$ & $1.68 \times 10^{-2}$ & 2.09 & $2.20 \times 10^{-1}$ & $4.26 \times 10^{-1}$ & 1.94 \\ 
1300 & $4.08 \times 10^{-4}$ & $8.86 \times 10^{-4}$ & 2.17 & $1.43 \times 10^{-3}$ & $3.03 \times 10^{-3}$ & 2.12 & $6.21 \times 10^{-2}$ & $1.21 \times 10^{-2}$ & 1.95 \\
1500 & $6.00 \times 10^{-5}$ & $1.36 \times 10^{-4}$ & 2.27 & $2.64 \times 10^{-4}$ & $5.73 \times 10^{-4}$ & 2.17 & $1.94 \times 10^{-2}$ & $3.82 \times 10^{-2}$ & 1.97 \\ 
\hline
\end{tabular}
\end{center}
\caption{Total cross sections $\sigma(pp \to e_8^+ e_8^-)$
and corresponding $K$ factors, for different
leptogluon masses and LHC energies. The renormalization and factorization scales
we fix to the central value $\mu_R = \mu_F = \mu^0 = m_{l_8}$.}
\label{tab:results}
\end{table}

We first concentrate on
the total cross section for leptogluon pair production, $\sigma(pp \to \lo\bar{l}_8)$,
computed to leading and next-to-leading order with the associated $K$ factors.
A comprehensive numerical survey 
we document 
in Table~\ref{tab:results}. We consider leptogluon masses ranging from 500~GeV to 1.5~TeV. 
From here on we shall concentrate on the production
of electron-octet pairs, $pp \to \eo\eobar$. The neutrino octet field $\nuo$
only contributes as a 
new colored virtual particle, 
entailing a mass-suppressed and hence numerically mild contribution. 
The constant ratio 
\begin{equation}
\frac{\sigma(pp \to \eo\eobar)}{\sigma(pp \to \nuo\nuobar)} = 2
\label{eq:ratio}
\end{equation}
reflects the fact that, so long as $m_{e^{\pm}_8} = m_{\nuo}$, 
the differences between the charged and the neutral channels reduce 
to a symmetry factor $1/2$, which accounts for the two identical particles
in the $\nuo\nuobar$ final state. The results we present for three
LHC nominal energies: $\sqrt{S}=7, 8, 14$~TeV. 

\bigskip{}

The total cross sections are sizable and very strongly dependent
on
the leptogluon mass. Leptogluons with $m_{\lo} \sim 500$~GeV would be typically produced
at a rate of $1-10$~pb.
TeV-scale leptogluons, in turn, range around
$\mathcal{O}(10)$~fb.

The size of the QCD quantum effects we quantify as 
$K\equiv \sigma^\text{NLO}/\sigma^\text{LO}$. This may attain very large values,
spanning the range of 1.8 to 2.0 for $\sqrt{S} = 14$~TeV, or even higher values 
2.0-2.2 for $\sqrt{S} = 8$~TeV. 
These uncomfortably large $K$-factors 
we should not interpret as a breakdown of perturbation theory.
Instead, they signal 
artificially low LO production
rates for $\mathcal{O}(1)$~TeV mass particles, a known
problem of the CTEQ parton densities~\cite{sgluon,susypair}.
The NLO cross sections are actually 
perturbatively stable.
On the other hand, 
the growth of the $K$ -factors as the leptogluon mass increases we can 
trace back in part to a non-trivial threshold behavior from the real and the virtual NLO
contributions, which partially overcomes the phase space suppression
of the NLO contributions. We will expand on all these aspects below.

\begin{figure}[t]
\includegraphics[width=0.35\textwidth]{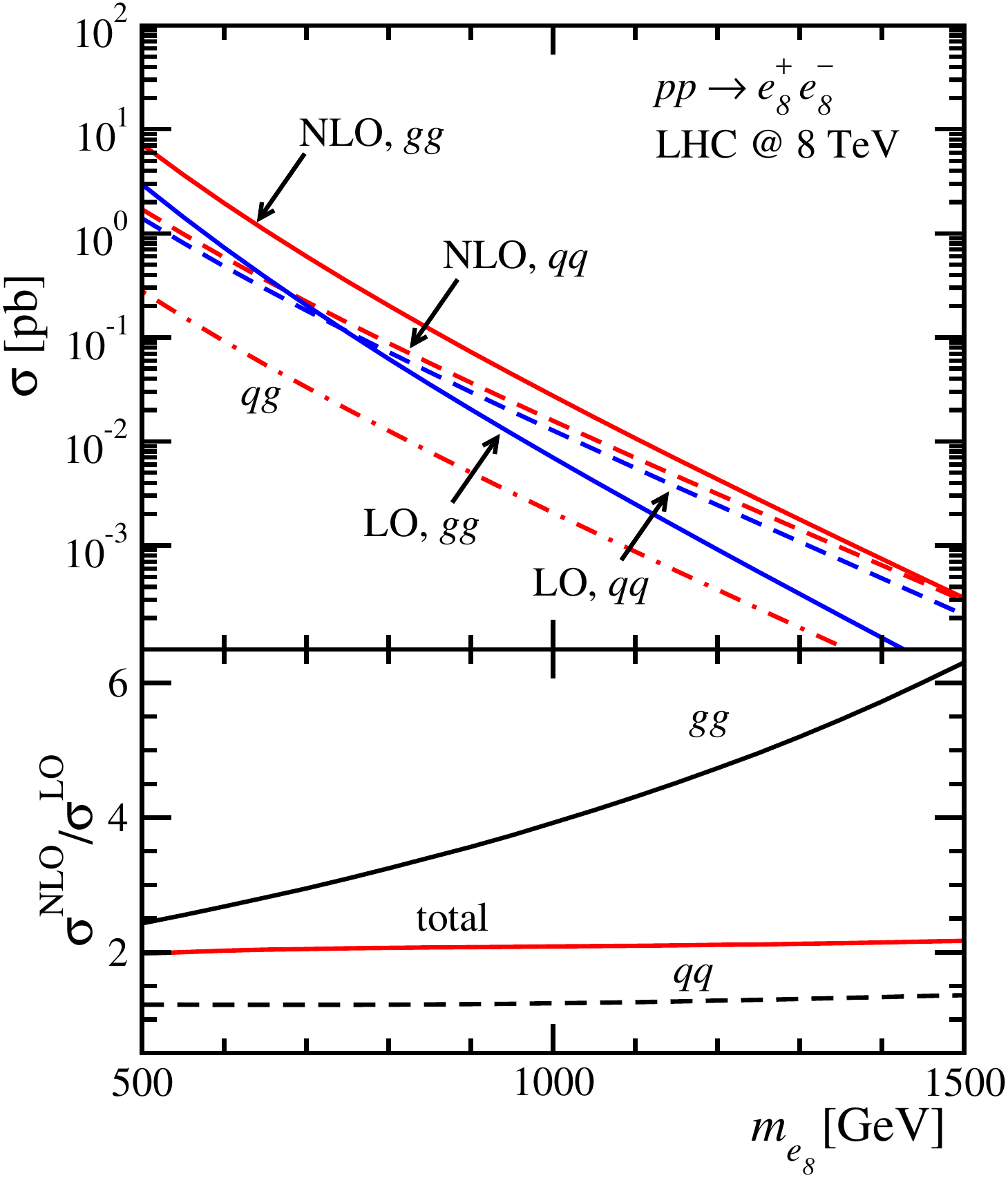}
\hspace*{0.1\textwidth}
\includegraphics[width=0.35\textwidth]{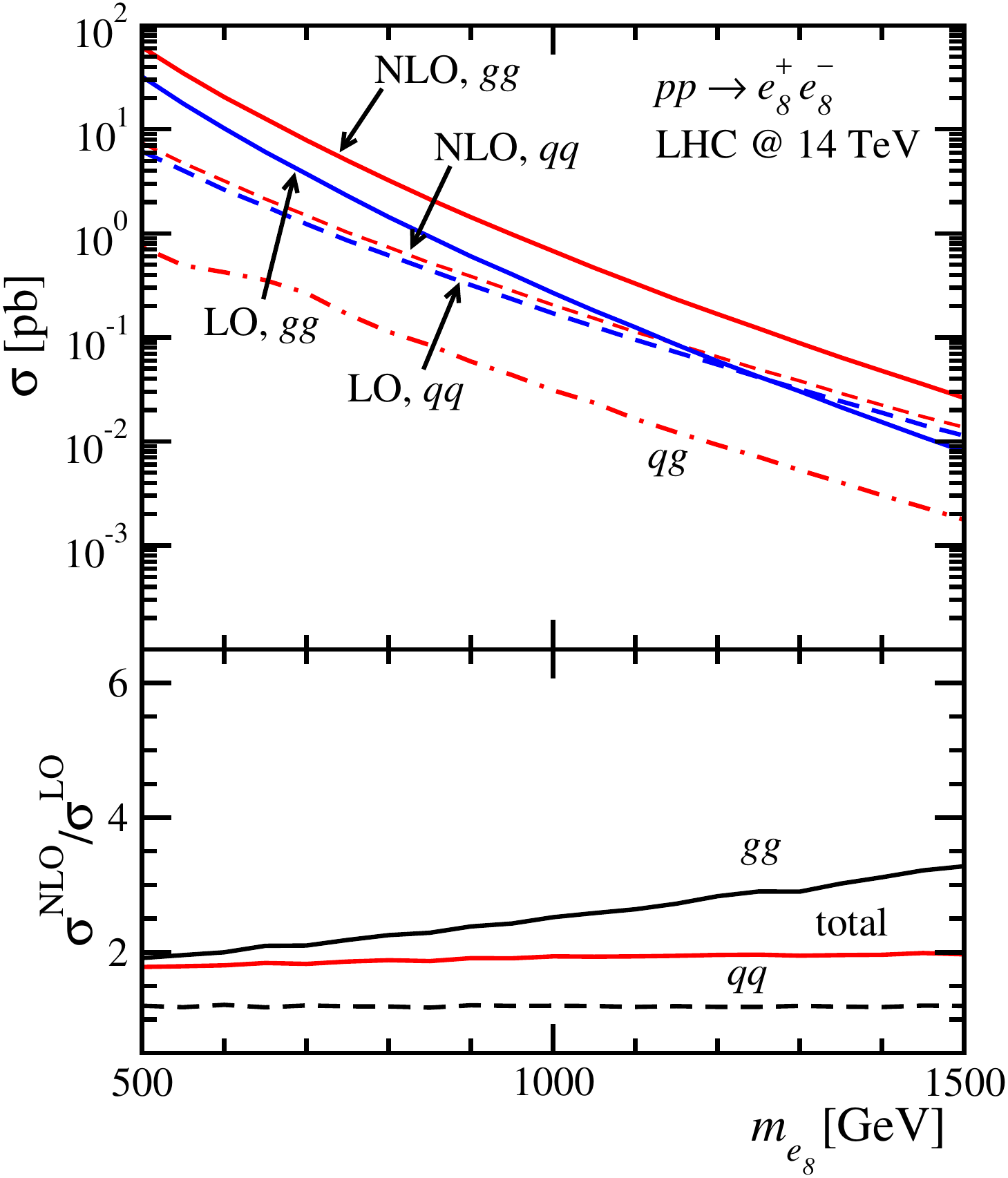} 
\caption{LO and NLO contributions 
from the $gg$ and $q\bar{q}$ partonic subchannels
to the 
total rates (above) and $K$ factor (below)
for leptogluon pair production $\sigma(pp \to \eo\eobar)$. 
The center-of-mass energy we fix at 8 (left panel) and 14~TeV (right).}
\label{fig:overmass}
\end{figure}

\bigskip{}
The total rates drop nearly three orders of magnitude when sweeping
the leptogluon mass range from 0.5 to 1.5~TeV. 
A closer view into this 
mass dependence we display
in Figure~\ref{fig:overmass}. Herewith we profile the cross section (top panel)
and $K$-factor (down panel) as a function of $m_{e_8^{\pm}}$ for the different partonic
subchannels. The LHC energy we fix to 8~TeV (left) and 14~TeV (right).
The dominance of the $gg$-initiated mechanism for low-mass leptogluons
--- roughly one order of magnitude above $q\bar{q}$ --- follows 
from the large gluon luminosity; for larger $m_{l_8}$ values this effect is inverted.
The $q\bar{q}$ mechanism includes not only
the contribution from the valence quarks $u,d$ but also the suppressed
flavor excitations from the second generation, which amount to less than 10\% of
the $q\bar{q} \to e^+_8\, e^-_8$ budget. The quark-gluon crossed 
channels $qg$, which arise purely at NLO, stagnate at the per-mil level given our treatment of the collinear divergences~\cite{susypair}.

\smallskip{}
In more detail, we can 
read off Figure~\ref{fig:overmass}
that $\sigma(gg \to \lo\lobar)$ to LO depletes faster than $\sigma(q\bar{q} \to \lo\lobar)$ with increasing 
leptogluon masses, \ie 
the associated $K$ factor
$K_{gg}$
exhibits a steeper growth. 
Unsurprisingly, we encounter a similar behaviour 
for gluino pairs~\cite{gluinopair}. 
These features we can understand, on the one hand, from the 
scaling 
at threshold, as we will explain in detail later. Moreover, they
can again be 
related to the respective
parton luminosities. Heavy leptogluon masses probe larger values
of the Bjorken $x$-variable -- this being the region where the quark parton densities 
become more competitive as compared to the gluon ones.

\bigskip{} 
It is illustrative
 to compare the leptogluon production rates
to similar QCD-driven heavy particle production, in particular
squarks~\cite{susypair,prospino_sqgl,prospino_stop}, gluinos~\cite{susypair,gluinopair}, and sgluons~\cite{sgluon}.
Based on 
{\sc MadGolem}~\cite{madgolem} and {\sc Prospino}~\cite{prospino_sqgl,prospino_stop}
we can establish the hierarchy
%
%
\begin{alignat}{5}
 \sigma(pp \to \lo\bar{l}_8) \sim \sigma(pp \to \go\go) \sim 
\mathcal{O}(10)\,\times\,\sigma(pp \to GG^*) \sim \mathcal{O}(100)\,\times\,\sigma(pp \to \tilde{q}\tilde{q}^*) \; .
 \label{eq:xsections}
 \end{alignat}
The differences we may track down to
larger color charges involved in the production of color-adjoints
as compared to fundamentals and to the spin representation
of the final-state particles.
The production of fermion pairs like leptogluons and gluinos is more efficient
than that of sgluons. Finally, it is also worth pointing out that the phenomenological
profile of leptogluon and leptoquark pair production is very similar, in spite of their
different spin and color representations. We will further exploit these coincidences in Section~\ref{sec:signatures}.

\subsection{Real emission}
\label{ssec:real}

\begin{figure}[t]
 \includegraphics[width=0.85\textwidth]{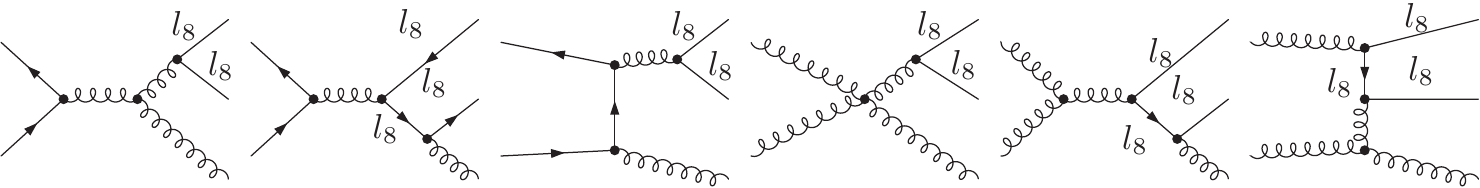} 
\caption{Sample Feynman diagrams describing the real emission
corrections for the pair production of leptogluons 
via quark-antiquark annihilation (left) and gluon fusion (right).}
\label{fig:real}
\end{figure}

\begin{figure}[b]
 \includegraphics[width=0.35\textwidth]{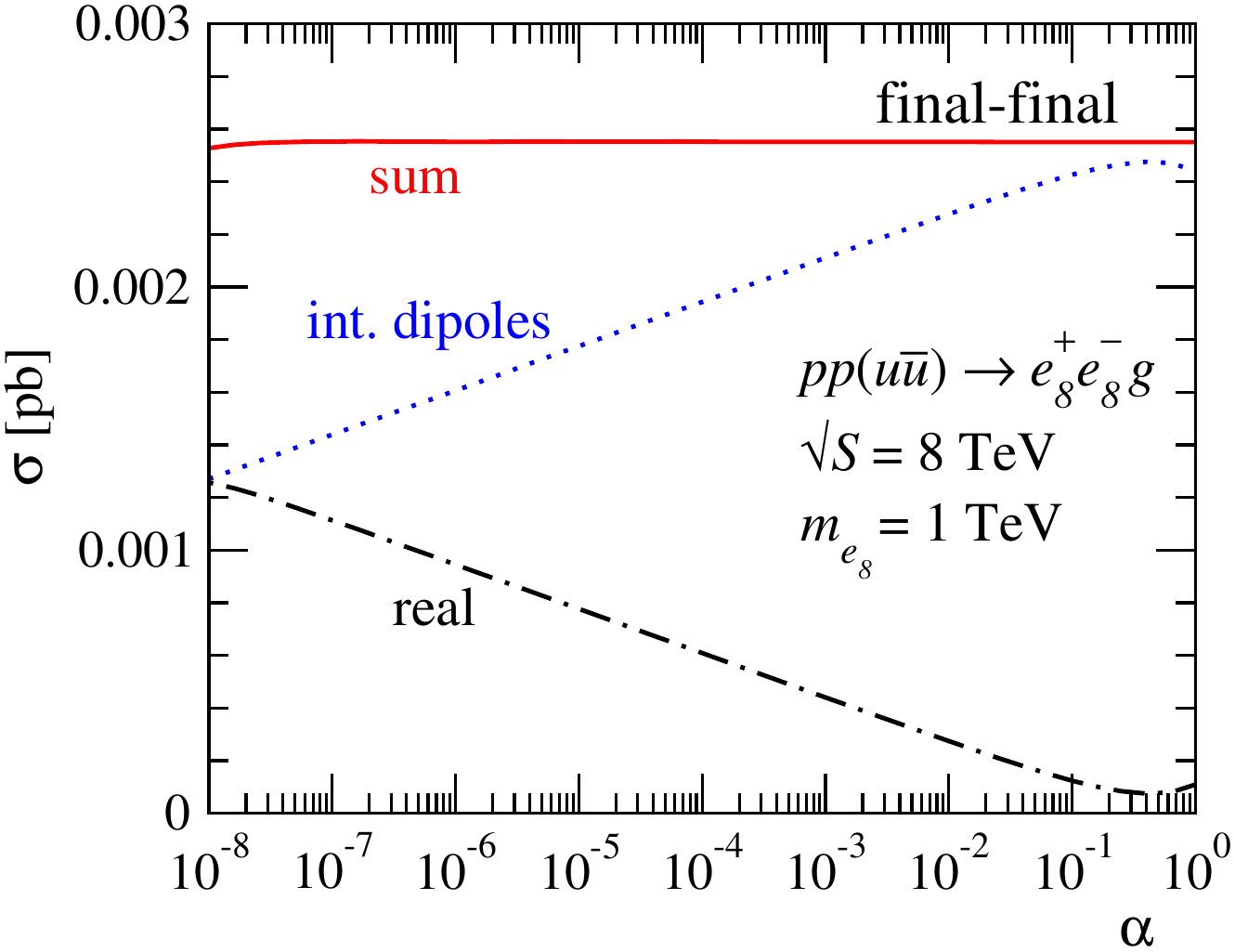} 
\caption{Dependence on the FKS-like $\alpha$ parameter corresponding
to the final-final leptogluon Catani--Seymour dipole for the partonic subchannel $u\bar{u} \to e_8^+e_8^- g$.
The leptogluon mass we fix at $m_{e^\pm_8} = 1$~TeV.}
\label{fig:dipole}
\end{figure}

Real emission is part of the QCD quantum effects to leptogluon pair production.
It comprises the $\mathcal{O}(\alpha_s^{3})$ three-particle final states with a leptogluon pair
plus an additional parton, which accounts for the QCD jet radiation,
either from the initial-state partons or the final-state leptogluons (cf. Figure~\ref{fig:real}). 

These contributions are
infrared divergent, as a result of the soft and collinear singularities developed by the initial-state
radiation, as well as the soft poles from the gluon radiation off the leptogluon legs.
These infrared divergences we subtract using massive Catani-Seymour dipoles~\cite{catani_seymour_massless,catani_seymour_massive}. On top of the Standard Model initial-initial and
initial-final dipoles \mg automatically provides the corresponding leptogluon dipoles which
are required for new final-final and final-initial singularities.
Since the leptogluons are color octet massive fermions, their dipoles are equivalent to the massive fermions with the
replacement of the color factors $C_F \to C_A$. 

The {\sc MadGolem} implementation is based
on an extended version of {\sc MadDipole}~\cite{maddipole} and 
retains a variable phase space coverage in terms of the
FKS-style cutoff $0<\alpha \leq 1$~\cite{alpha}. As the default value we use $\alpha=1$.
The numerical performance of the leptogluon dipole we can assess from Figure~\ref{fig:dipole}.
Notice that while the individual contributions from the integrated and the non-integrated
dipoles diverge logarithmically for small $\alpha$ their sum is stable
over roughly eight orders of magnitude. These results we compute for the 
$u\bar{u} \to e_8^+e_8^- g$ partonic subprocesses, assuming 1~TeV leptogluon final states.

\subsection{Virtual corrections}
\label{ssec:virtual}

\begin{figure}[t]
 \includegraphics[width=0.85\textwidth]{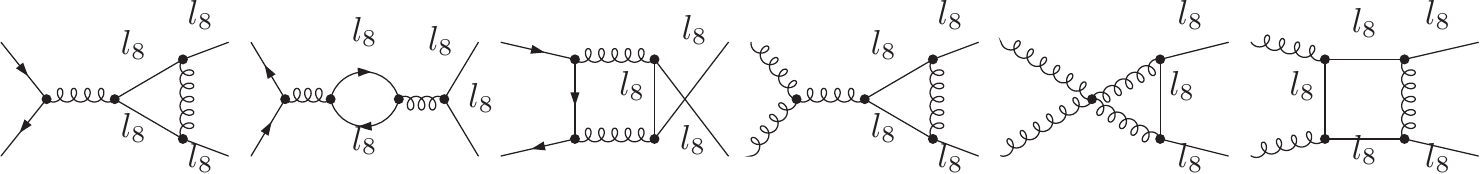} 
\caption{Sample Feynman diagrams describing the NLO virtual
corrections to the pair production of leptogluons 
via quark-antiquark annihilation (left) and gluon fusion (right).}
\label{fig:virtual}
\end{figure}

\begin{figure}[b]
\includegraphics[width=0.38\textwidth]{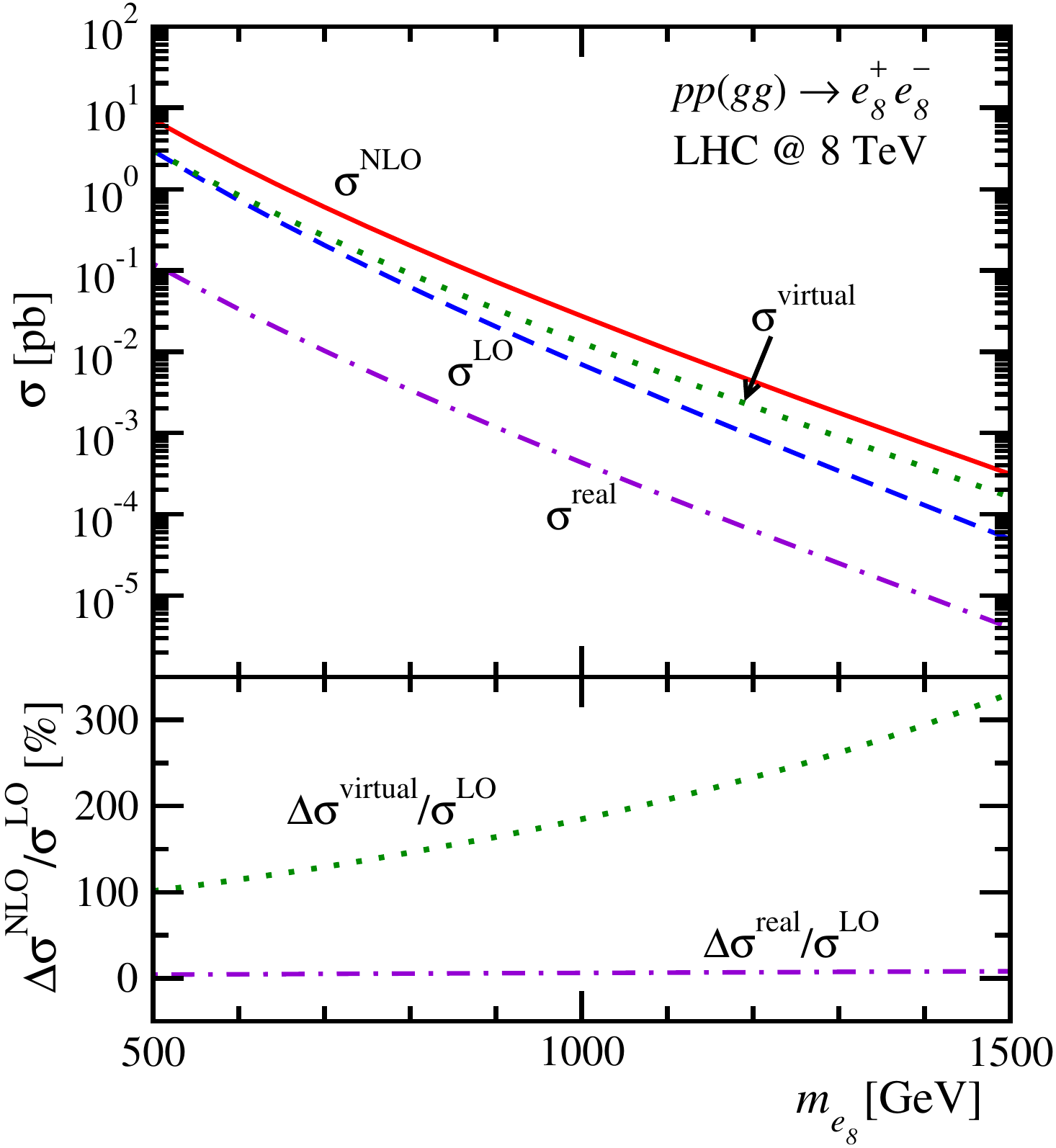}
\hspace*{0.1\textwidth}
\includegraphics[width=0.38\textwidth]{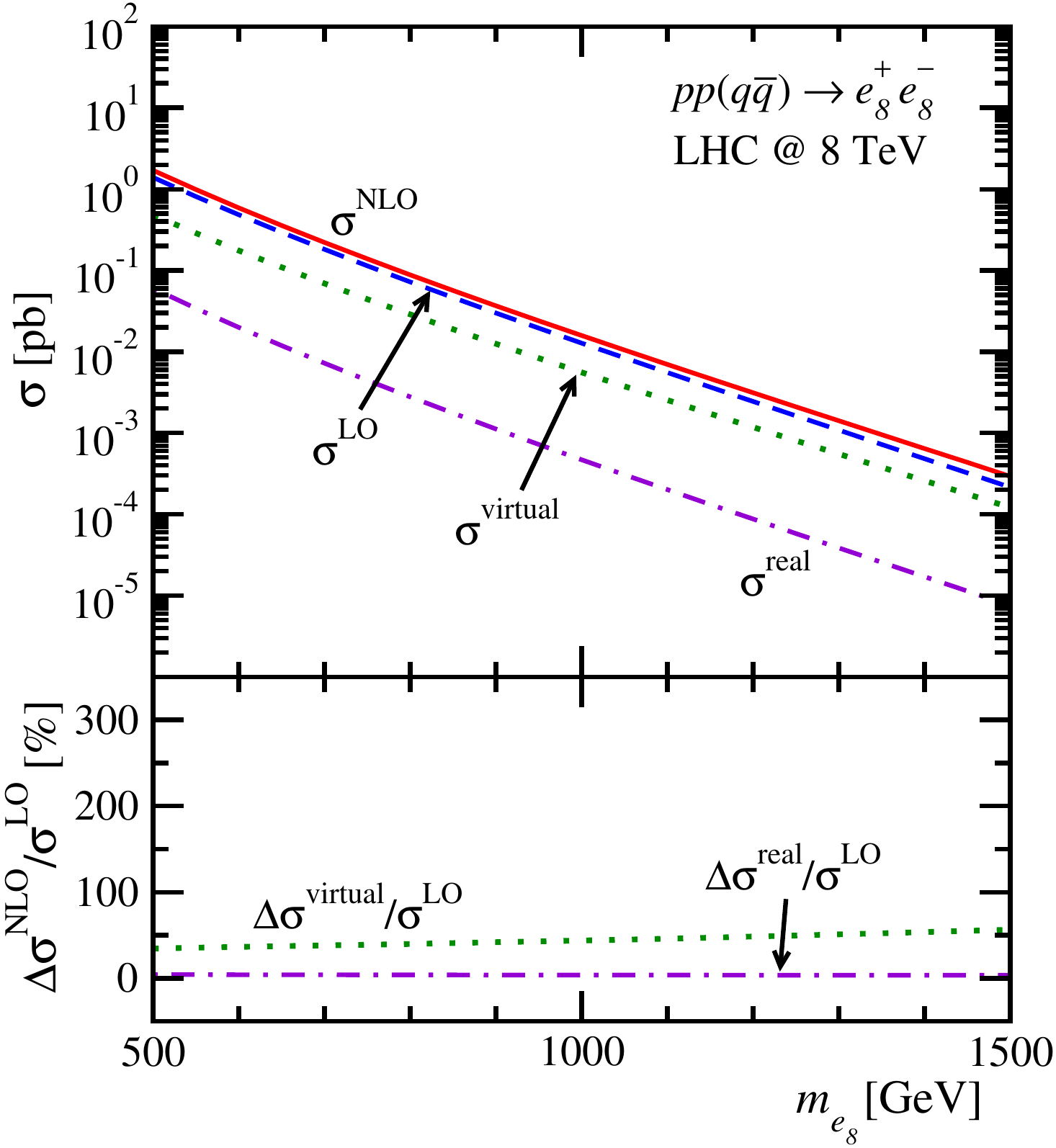}
\caption{Total LO and NLO leptogluon pair cross sections $\sigma(pp \to \eo\eobar)$ 
as a function of the mass.
We separate 
gluon fusion (left)
and 
quark/antiquark annihilation (right).
In the lower 
panels we account for the relative size of the 
corrections, separated 
using Catani--Seymour dipoles with $\alpha=1$ and including the integrated
dipoles as
part of the virtual contribution.}
\label{fig:virtuala}
\end{figure}

The second class of NLO quantum effects includes $\mathcal{O}(\alpha^3_s)$ terms 
from the exchange of virtual gluons, quarks and leptogluons, 
giving rise to self-energy, vertex and box one-loop corrections. A sample of the corresponding
Feynman diagrams we display in Figure~\ref{fig:virtual}. Again, we compute them
using {\sc MadGolem}~\cite{madgolem}, generating
the one-loop amplitudes 
with {\sc Qgraf}\cite{qgraf}. To handle the helicity and color structures we 
rely on a dedicated set of currently in-house routines, prior to finally reducing their tensor structures in the {\sc Golem}~\cite{golem,golem_lib} framework.
The resulting one-loop integrals we numerically evaluate with the {\sc OneLoop} library~\cite{oneloop}.  
For a detailed account on the structure and performance of the tool, we refer the reader to Ref.~\cite{madgolem}.

\smallskip{}
All divergences we regularize using standard 
dimensional regularization, 
analytically extending
all integrals and internal propagators
to $n = 4-2\epsilon$ dimensions. We then subtract the infrared poles 
by including the
integrated Catani--Seymour dipoles. The collinear higher-order corrections we absorbe
into the parton densities. The ultraviolet divergences in turn cancel against
the corresponding ultraviolet counter terms, which are part of the \emph{leptogluon\_nlo} 
model implementation and are generated automatically. Its structure is completely defined by 
the on-shell renormalization of 
the leptogluon mass and the 
\msbar\, renormalization of the strong coupling constant,
with a decoupling of the heavy colored degrees of freedom~\cite{decoupling}.
We check the cancellation of divergences both analytically and numerically. The finite
parts for the one-loop amplitude we compare to an independent implementation
of the leptogluon model in {\sc FeynArts}, {\sc FormCalc} and {\sc LoopTools}~\cite{feynarts}.
In the Appendix we supply further details on the renormalization.

\begin{figure}[t]
\includegraphics[width=0.4\textwidth]{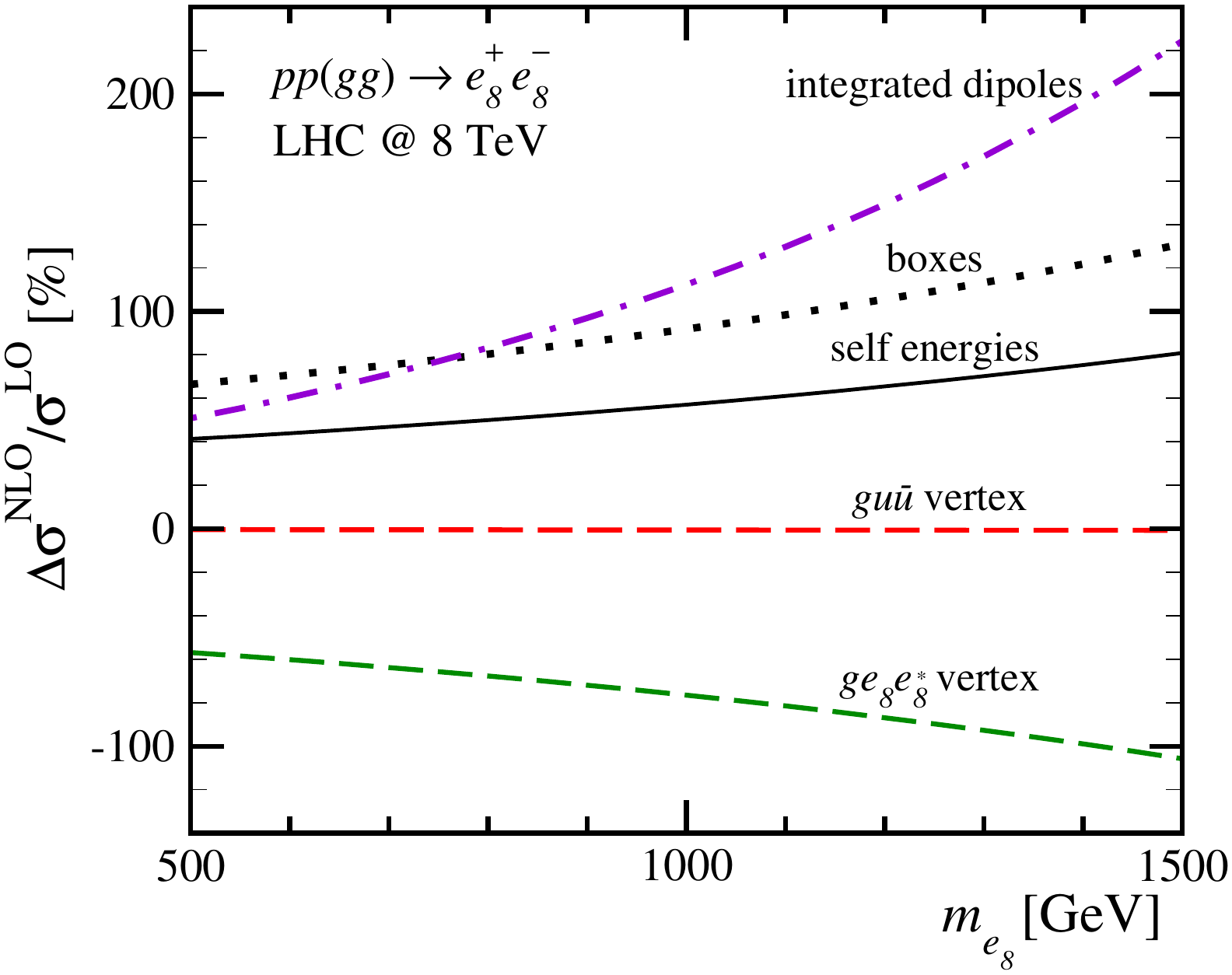}
\hspace*{0.05\textwidth}
\includegraphics[width=0.4\textwidth]{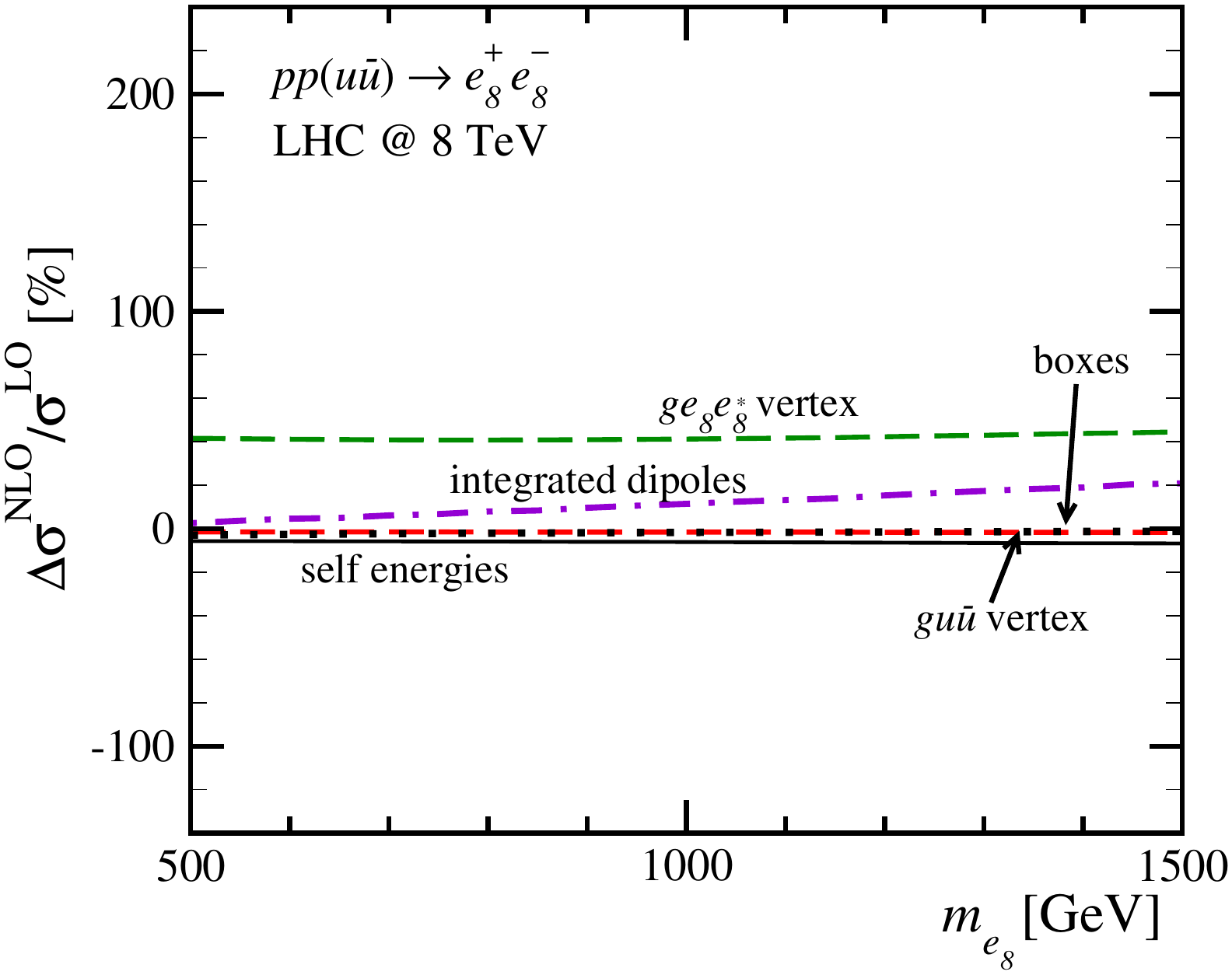}
\caption{Relative size $\Delta \sigma^\text{NLO}/\sigma^\text{LO} \equiv 
(\sigma^\text{NLO} - \sigma^\text{LO})/\sigma^\text{LO}$
for the different
one-loop contributions
as a function of the leptogluon mass.
The two leading-order parton channels $\sigma(gg\to\lo\lo)$ (left)
and $\sigma(u\bar{u} \to \lo\lo)$ (right) we show separately. 
The integrated dipoles we show for $\alpha = 1$.}
\label{fig:virtualb}
\end{figure}

\bigskip{}
We examine in detail the anatomy of the virtual corrections 
in Figures~\ref{fig:virtuala}-\ref{fig:virtualb}.
The patterns 
share many common features with sgluon~\cite{sgluon} and gluino~\cite{susypair,gluinopair} pair production.
To start with, in Figure~\ref{fig:virtuala} we 
survey the mass dependence of the real and the virtual corrections
for both parton channels $gg$ and $q\bar{q}$. For each of them we 
superimpose the total LO and NLO rates, together with the partial NLO virtual and real contributions
$\sigma^\text{virtual}$ and $\sigma^\text{real}$ (upper panels), while the relative size
of the NLO corrections 
we depict below. 

\smallskip{} 
The different subsets of one-loop topologies we examine in Figure~\ref{fig:virtualb}.
Their relative weight with respect to 
the LO results we describe via the
ratio
$\Delta\sigma/\sigma^\text{LO} \equiv (\sigma^{{\rm NLO},i}-\sigma^\text{LO})/\sigma^\text{LO}$,
where $i$ runs over all the different 
one-loop contributions --- including gluon emission which enters 
through the integrated dipoles. Again, we separate 
the $gg$ and $q\bar{q}$ subchannels. The crossed channel $qg$ does not develop
any virtual corrections and hence it is not included.
Notice, though, that
it is required to achieve a complete cancellation of the collinear divergences.
The figure unveils, first of all,
large vertex corrections to the QCD coupling $g e_8^+ e_8^-$ 
of at least $\mathcal{O}(40 \%)$ and with opposite signs for each
of the subchannels. 
The corresponding mass dependence turns out to be relatively mild, 
with a total variation not larger than $\sim 10\%$ for the whole leptogluon mass range
under scrutiny.

Sizable (positive) corrections are driven by
the integrated Catani--Seymour dipoles, \ie 
from soft/collinear gluon emission. 
Also large are the box contributions for the $gg$ initial state. 
Finally, 
 leptogluon self-energy corrections become large for the $gg$ initial state. 
 Interestingly, they all feature a characteristic growth 
with the leptogluon mass.
Both, the size and the acute mass dependence of these
subsets we can trace back to genuine kinematical and dynamical features.

\begin{figure}[b]
\includegraphics[width=0.4\textwidth]{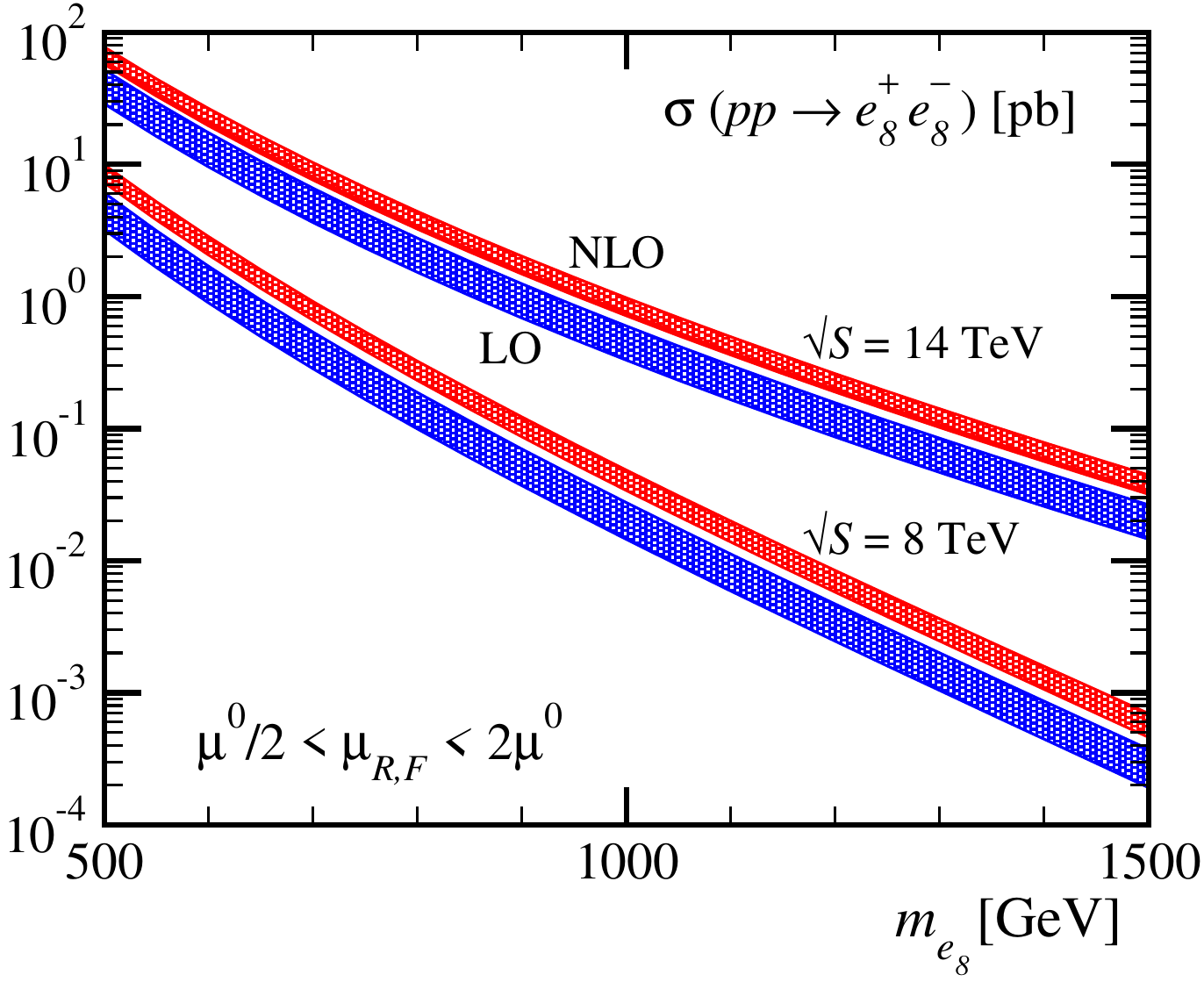}
\caption{Leptogluon pair cross section $\sigma(pp \to \eo\eobar)$ as a
  function of the leptogluon mass. The scale uncertainty envelope
  corresponds to an independent renormalization and factorization scale variation
  in the range $\mu^0/2 < \mu_{R,F}< 2\mu^0$.}
\label{fig:overmass-band}
\end{figure}

The steeper growth of the $K$ factors for the $gg$-initiated 
subprocess can
be in part understood from the LO threshold behavior:
there, the rate scales like
$\sigma_{gg} \sim \pi\,m^2_{l_8}/\hat{S}$
versus 
$\sigma_{q\bar{q}} \sim \pi\,m^2_{l_8}\beta/\hat{S}$,
with $\beta^2 = 1-4m^2_{l_8}/\hat{S}$, which means that
$\sigma^\text{LO}(gg \to \lo\lo)$ decreases faster
than $\sigma^\text{LO}(qq \to \lo\lo)$ with growing leptogluon mass.
The dominant virtual corrections benefit from two threshold effects.
An enhanced
Sommerfeld rescattering~\cite{sommerfeld} describes the exchange of
soft gluons between slowly-moving final-state leptogluons, leading 
to a Coulomb singularity $\sigma^\text{virtual} \sim \pi\alpha_s/\beta$.
The complete logarithmic enhancement of the soft gluon emission behaves like
$\sigma \sim A\log^2(\beta) + B\log(8\beta^2)$. 

Both sources of threshold enhancements
become more
efficient close to threshold for the steep gluon luminosities and at large leptogluon masses,
and can eventually be resummed~\cite{threshold}.

The enhanced leptogluon self energies in the $gg \to \lo\lobar$ case
can be correlated to the leptogluon mass insertion in
the $t$ and $u$-channel 
leptogluon exchange -- with
no counterpart for $q\bar{q} \to \lo\lo$.
The remaining one-loop contributions, consisting of 
QCD-like quark-gluon
and triple-gluon vertex corrections and the box and self-energy corrections 
for $q\bar{q} \to \lo\lo$ are essentially featureless and contribute 
with
overall yields below 10\%.

\subsection{Scale dependence}
\label{ssec:scale}

One of the main motivations of higher-order predictions is to stabilize the
(unphysical) dependence on the renormalization and factorization
scales. We introduce them when we remove the collinear and ultraviolet
divergences at a given order in the perturbative series. In the limit where all terms
in such series are retained the scale dependence vanishes.
In practice,
we truncate the expansion on the coupling constant
at a certain order $\mathcal{O}(\alpha_s^n)$. 
In the absence of any systematic correlation between different orders $n$ 
we expect the scale dependence to englobe the asymptotic
cross-section values. This expectation has been well confirmed for many 
QCD-mediated
processes~\cite{sgluon,susypair,prospino_sqgl,prospino_stop}.\smallskip

In Figure~\ref{fig:overmass-band} we depict the 
evolution of the 
total leptogluon pair cross sections at LO and NLO 
with the leptogluon mass, for the LHC operating at 8 and 14~TeV. 
The band envelope accounts for the theoretical uncertainty,
which we estimate as conventionally by varying the scale choices within the range 
$\mu^0/2 < \mu_{R,F} < 2\mu^{0}$. 
We see that over the entire mass range the LO and NLO curves hardly overlap, which means that at least for the LO cross section a scale variation 
by a factor two around the central scale would not have been a conservative estimate of the theory uncertainty due to missing higher orders in perturbation theory.

\begin{figure}[t]
\includegraphics[width=0.8\textwidth]{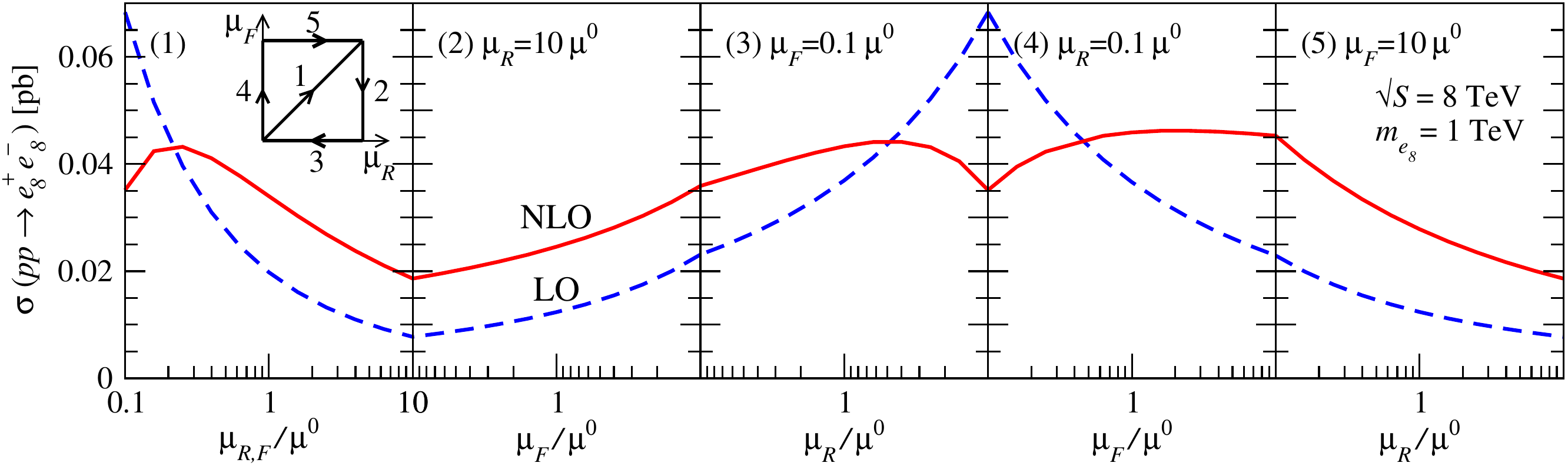}
\caption{Renormalization and factorization scale dependence for leptogluon
  pair production $pp \to e_8^+ e^-_8$. 
  The plots trace a contour in the $\mu_R$-$\mu_F$ plane in
  the range $\mu = (0.1 - 10) \times \mu^0$ with $\mu^0 = m_{e^{\pm}_8}$, 
  and $m_{e^{\pm}_8} = 1$~TeV.}
\label{fig:scales-running}
\end{figure}

\smallskip
Complementary vistas on the stabilization of the scale dependence we provide
in Figure~\ref{fig:scales-running}. We delineate the LO and NLO cross sections
for an independent variation of the factorization
and renormalization scales. The five panels correspond to a contour in the 
$\mu_R-\mu_F$ plane, which we define in the left panel.
The smoothing of $\sigma(\mu)$ at NLO nicely illustrates the 
stabilization
of the higher order prediction with respect to scale choices.
Quantitatively, the LO variation stays within 
$\Delta \sigma^\text{LO}/\sigma^\text{LO} \sim \mathcal{O}(65\%)$, while at NLO
it reduces to $\Delta\sigma^\text{NLO}/\sigma^\text{NLO} \sim\mathcal{O}(30\%)$.

\subsection{Distributions vs jet merging}
\label{ssec:distrib}

Relying on fixed order NLO predictions we 
can extract 
a suitable normalization to the event 
rates from standard Monte Carlo simulations. Nevertheless, one still 
needs to assure that this framework can be safely promoted to the main distributions. 
It has been shown in the literature
that transverse momentum and rapidity distributions are relatively stable with
respect to higher order corrections for heavy particle production ~\cite{sgluon,susypair,prospino_sqgl,sqn1}.
As long as 
the 
collinear approximation includes sizable $p_T$ values owing to even larger final-state
masses
QCD jet radiation
should be properly accounted for by the parton shower description~\cite{skands,sgluon,madgraph_merging}.\bigskip

\begin{figure}[h!]
\includegraphics[width=0.455\textwidth]{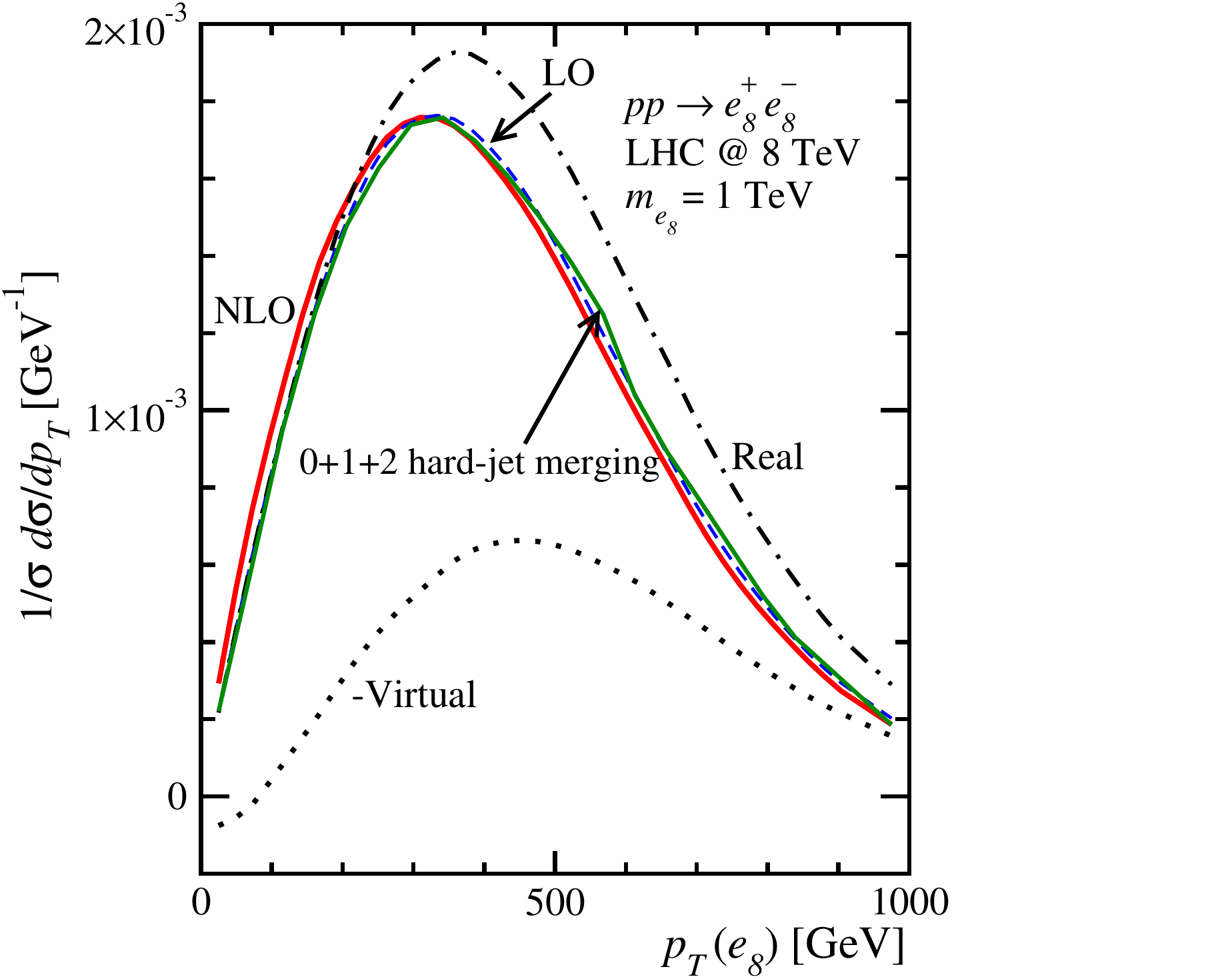}
\includegraphics[width=0.315\textwidth]{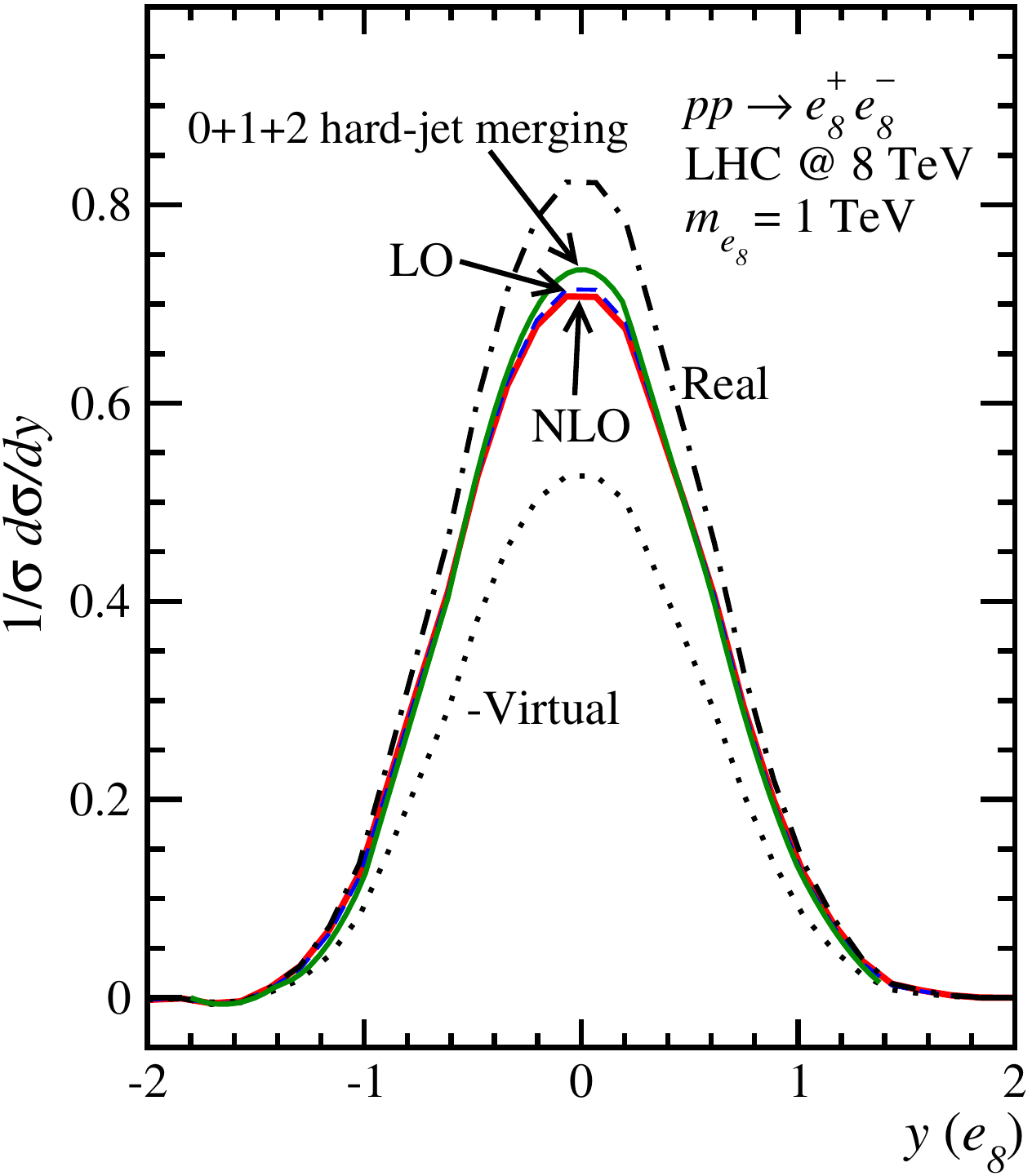}
\caption{NLO distributions for leptogluon pair production as a
 function of the transverse momentum $p_T(e^{\pm}_8)$ and rapidity $y(e^{\pm}_8)$. 
  For the NLO curves we separately display the LO, virtual, and real
  contributions, separated using 
  Catani--Seymour dipoles with $\alpha = 10^{-3}$. In addition, we
  show the distributions based on LO matched samples with up to two
  hard jets. The NLO and matched curves we normalize to unity while
  the different contributions to the NLO rates we show to scale.}
\label{fig:distrib}
\end{figure}

In Figure~\ref{fig:distrib} we 
check
these expectations. Therewith, we display the transverse momentum and rapidity distributions
for 1~TeV leptogluons. The different LO and NLO pieces we display separately.
Real emission and virtual NLO effects we disentangle by introducing Catani--Seymour dipoles
and separating the corresponding phase space regions within $\alpha = 10^{-3}$.
We can then compare the fixed-order NLO results from \mg with a matched matrix element and parton shower description 
using the MLM scheme including up to two hard jets~\cite{mlm,lecture}. 
For the latter we employ {\sc MadGraph5}~\cite{mg5,madgraph_merging}
interfaced with {\sc Pythia}~\cite{pythia}. 
We explicitly check that including just one hard jet barely changes the results, 
in line with the general statement that in our case the collinear parton shower is not a bad approximation.

The two main histograms
we normalize to unity, with the different NLO contributions shown to scale. 
Both approaches display very similar
shapes, featuring central leptogluons with transverse momentum peaking
at $p_T(e_8^{\pm}) \simeq 300$~GeV.
Only when analysed individually,
the $p_T$ curves for the real and virtual NLO corrections
depart significantly from the combined curve.
The observed shift
towards slightly harder and more central leptogluons in the merged
result, which also appears for 
sgluon pairs~\cite{sgluon}, we can deem to the additional recoil jets.

\begin{figure}[bth]
\includegraphics[width=0.315\textwidth]{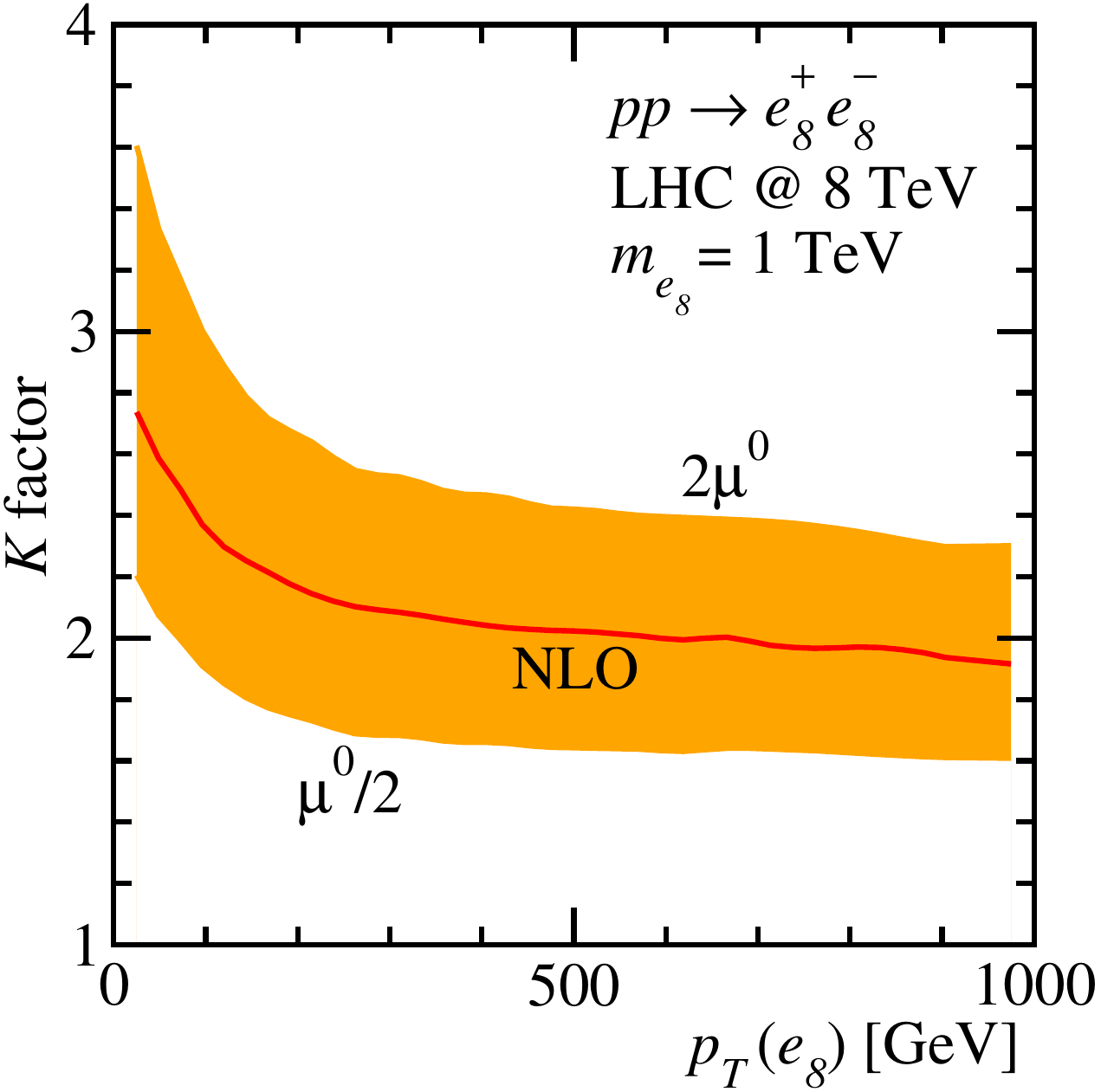} 
\hspace*{0.1\textwidth}
\includegraphics[width=0.3\textwidth]{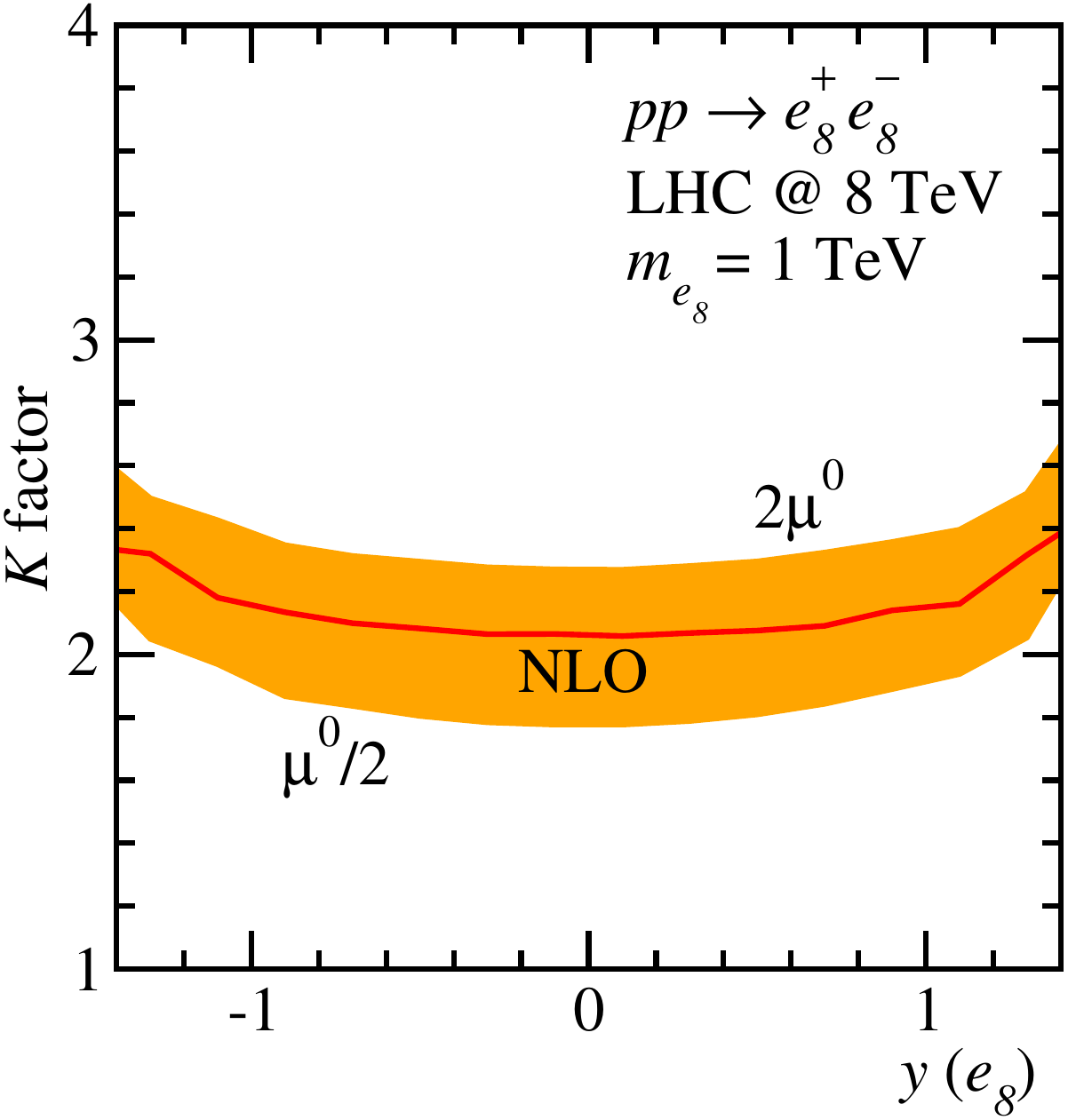}
\caption{$K$ factor as a function of the leptogluon transverse momentum $p_T(e_8^{\pm})$
and rapidity $y(e_8^{\pm})$.
We include the scale uncertainty envelope for an independent 
renormalization and factorization scale variation
within $\mu^0/2 < \mu_{R,F}< 2\mu^0$.}
\label{fig:distrib-band}
\end{figure}

\smallskip{}
Finally, in Figure~\ref{fig:distrib-band} we present a panoply of Figure~\ref{fig:distrib}, 
now displaying the transverse momentum and rapidity dependence
of the $K$-factors including the scale dependence envelope around the
central scale $\mu^0 = m_{e_8^{\pm}}$. Remarkably, the $K$-factors remain 
relatively constant within the remaining NLO scale dependence as long as
we stay within 
central rapidities and $p_T$ values around $m_{e_8^{\pm}}/2 \gtrsim 500$~GeV. This justifies
the use of the $K$-factor as an overall reweighting factor for the leading-order event
samples as long as we do not probe extreme phase space regions -- see \emph{e.g.} \cite{Dreiner:2012sh} for a counter example.

\section{Leptogluon signatures}
\label{sec:signatures}

\begin{figure}[t]
 \includegraphics[width=0.4\textwidth]{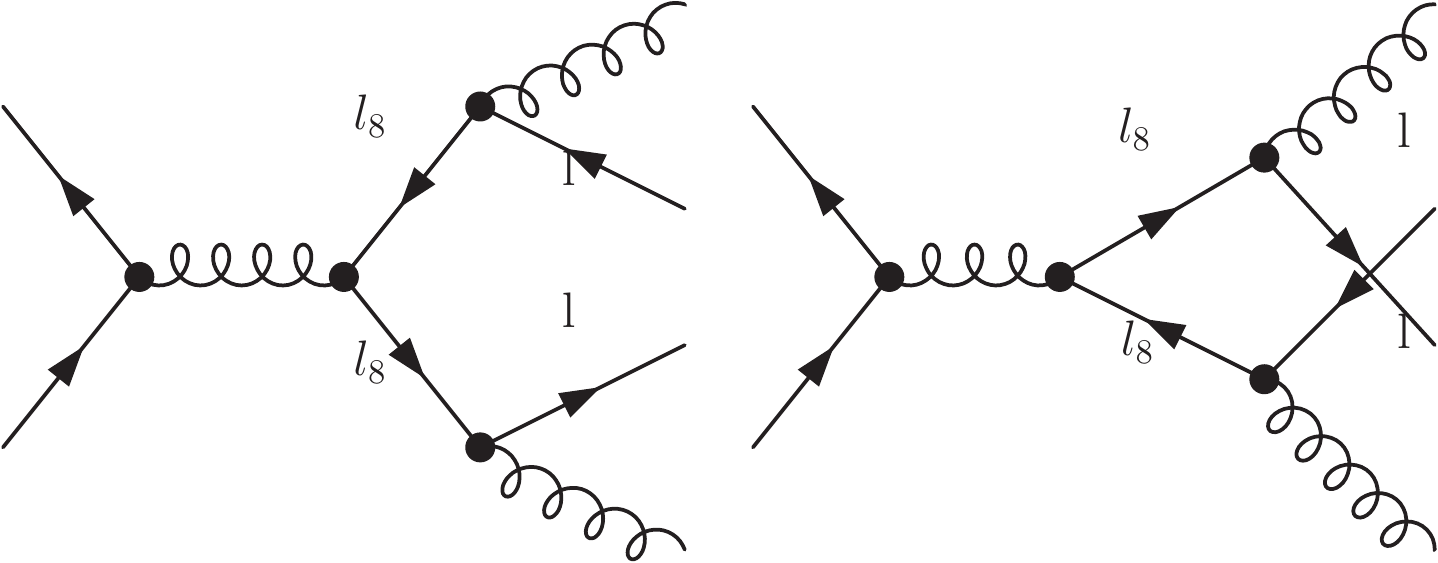} 
\caption{Sample Feynman diagrams describing the characteristic dijet--dilepton
final-state from leptogluon pair production
followed by a radiative transition into a Standard Model lepton and a gluon each.}
\label{fig:2to4}
\end{figure}

From here on we will focus on
trademark collider signatures
of leptogluon pairs and discuss them
in the light of the current LHC searches. Specifically,
we discuss how 
to make use of ongoing CMS searches for leptoquarks~\cite{:2012dnb}
to provide updated leptogluon mass bounds.

\bigskip{}

Besides the $SU(3)_c$ gauge couplings
leptogluons can also interact with 
matter fields.
These interactions are ultimately
induced by the dynamics of the underlying model.
They
appear as non-renormalizable couplings which we can parameterize
through higher dimensional effective operators. For simplicity, we 
limit ourselves to the dimension-5 chromomagnetic 
operator. 
It is a representative
example of such non-renormalizable couplings and accounts for the radiative decay
of a leptogluon decay into a lepton and a gluon.
This process is governed by the effective Lagrangian
\begin{equation}
\lag = \frac{g_s}{2\Lambda} \;
       \overline{\psi}_{l_8}^A \; \sigma^{\mu\nu}\, G^A_{\mu\nu} 
           (a_L\,P_L + a_R\,P_R)\; \psi_l \, + \text{h.c.}  \; ,
\label{eq:hdim}
\end{equation}
with $\Lambda$ denoting the characteristic scale of the 
ultraviolet completion, \eg the compositeness
scale of a hypothetical preonic model.
The spinors $\psi^A_{l_8}$ and $\psi_l$ describe the leptogluon and Standard Model lepton.
In the presence of the gluon we factorize the strong coupling constant
$g_s$. The overall coupling strength is affected 
by the parameters $a_{L/R}$ which we expect to be of $\mathcal{O}(1)$.
If instead of a strongly-interacting ultraviolet completion
this transition is mediated by quantum effects we expect 
another suppression factor 
$1/16\pi^2$. Henceforth,
we assume $a_R= 0$, in agreement with electron chirality conservation
and the left-handed nature of the Standard Model neutrinos.

\medskip{} 
In the decay of leptogluons into a Standard Model lepton and a gluon
both (massless) daughter particles emerge as back-to-back
hard objects in the leptogluon rest frame.
The leptogluon partial width we compute
from the effective Lagrangian Eq.\eqref{eq:hdim}, and we may cast it into the following form 
\begin{eqnarray}
 \Gamma_{l_8 \to l q} = \frac{g_s^2\,m_{l_8}^3}{16\pi\Lambda^2}\,(a_L^2+a_R^2)
\label{eq:decay}.
\end{eqnarray}
This form 
coincides with the width of a gluino decaying into a gravitino-gluon
final state~\cite{Ambrosanio:1996jn} after matching the respective
effective couplings.  Such analogies between leptogluons and gluinos
are also rubber-stamped into their respective collider footprints,
cf. Ref~\cite{deAquino:2012ru} for a detailed account on associated
gluino-gravitino production at the LHC.\bigskip

We 
show a sample
of leading-order Feynman diagrams for the pair production of
leptogluons and their subsequent radiative decay
in Figure~\ref{fig:2to4}. 
The overall rates we can describe within the narrow width approximation, 
\begin{equation}
 \sigma(pp \to l_8 \bar{l}_8 \to l l g g) \simeq \sigma(pp \to l_8 \bar{l}_8)\times [\text{BR}(l_8 \to lg)]^2 \; .
\label{eq:narrow}
\end{equation}
\noindent The narrow width approximation is manifest from 
Eq.\eqref{eq:decay}
and assuming $a^2_{L(R)}/\Lambda^2 \ll 1$.
As a 
consequence
the 
leptogluon signatures will essentially not depend on the effective
higher-dimensional interactions modelling the leptogluon decay. 
We should be able to derive universal constraints on the leptogluon masses
unrelated to the characteristic ultraviolet scale $\Lambda$ and the coupling strength $a_{L(R)}$ of the
effective $l_8 l g$ interactions.

\bigskip{} As already mentioned, searches for
(charged) leptogluons at the LHC should 
close follow searches 
for leptoquarks, as both processes share the same final-state signature~\cite{:2012dnb}.
It consists of two isolated hard leptons
alongside two central jets. These leptons get replaced by neutrinos or 
missing energy when we instead entertain the case of neutral leptogluons.
As heavy resonances, leptogluon decays naturally give rise
to hard $p_T$ profiles for the resulting products, as well as sizable
invariant masses for the dilepton $m(e^+ e^-)$ and 
leading-$p_T$ dijet systems $m(jj)$.  
Irreducible Standard Model backgrounds 
are $Z/\gamma^*/W + $jets,
$W^+ W^- + $jets, and 
$t\bar{t}$ production. Additional subleading sources are
single top production, electroweak dibosons, 
and QCD multijets, typically well below the per-cent level.

\begin{figure}[t!]
\includegraphics[width=0.35\textwidth]{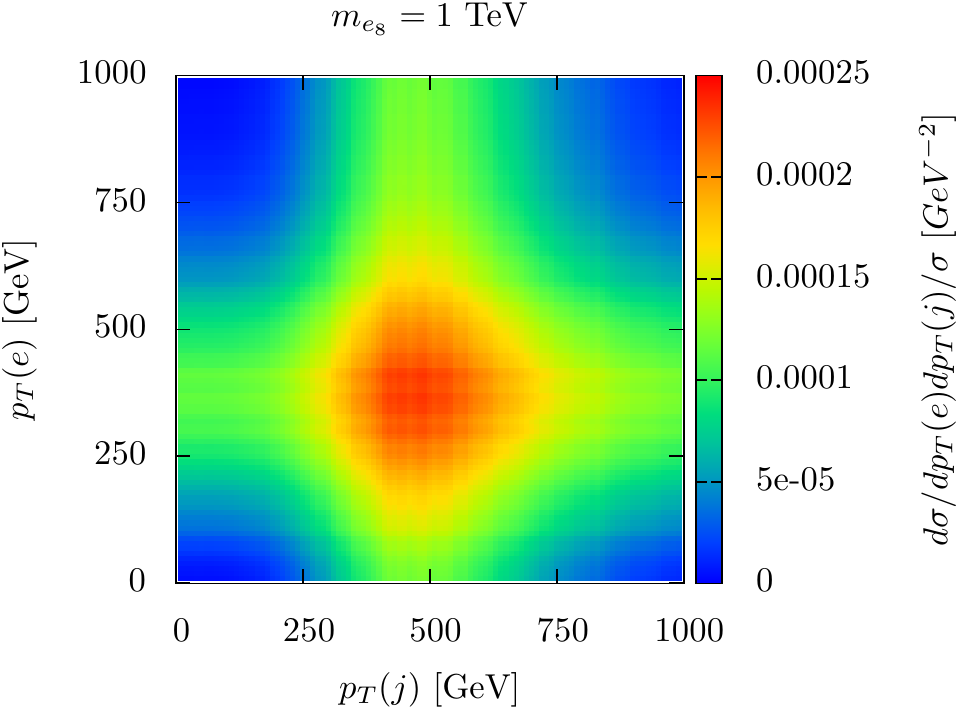}
\hspace*{0.1\textwidth}
\includegraphics[width=0.35\textwidth]{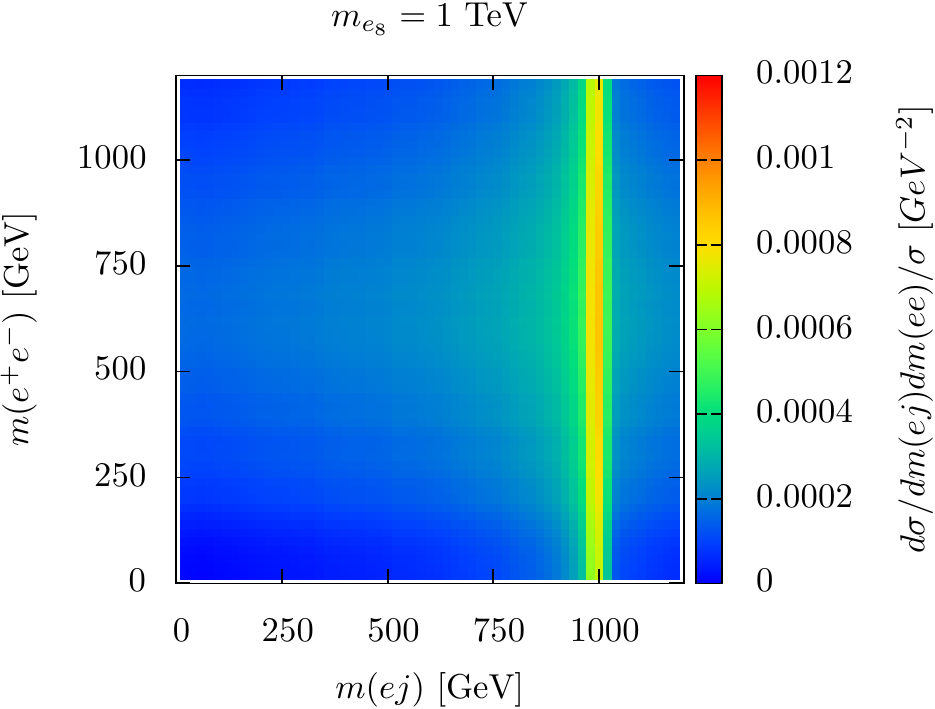} \\
\includegraphics[width=0.35\textwidth]{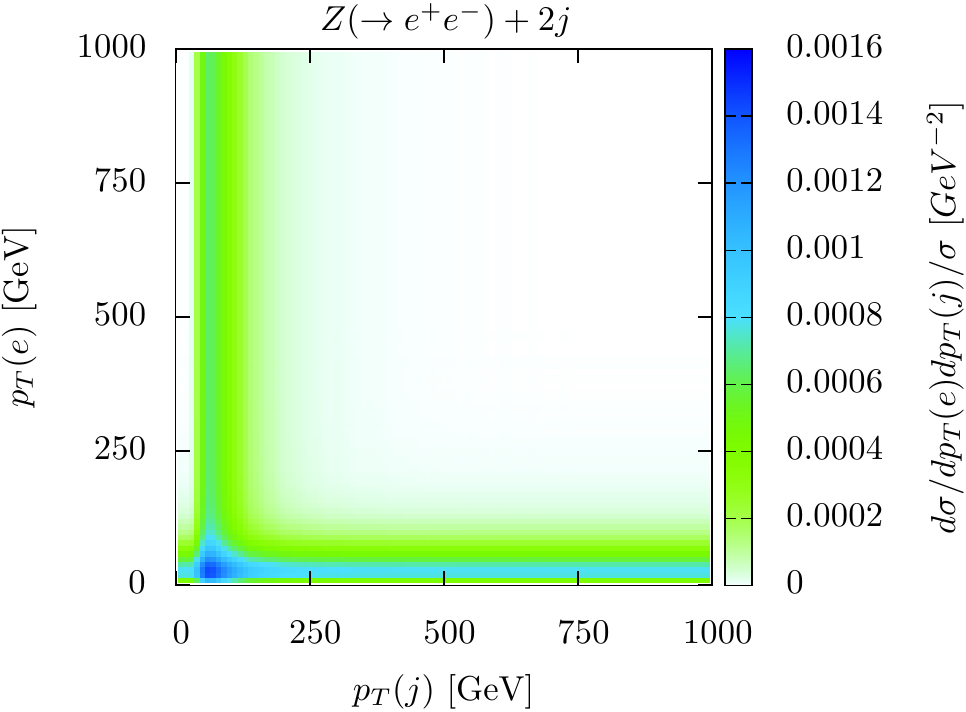}
\hspace*{0.1\textwidth}
\includegraphics[width=0.35\textwidth]{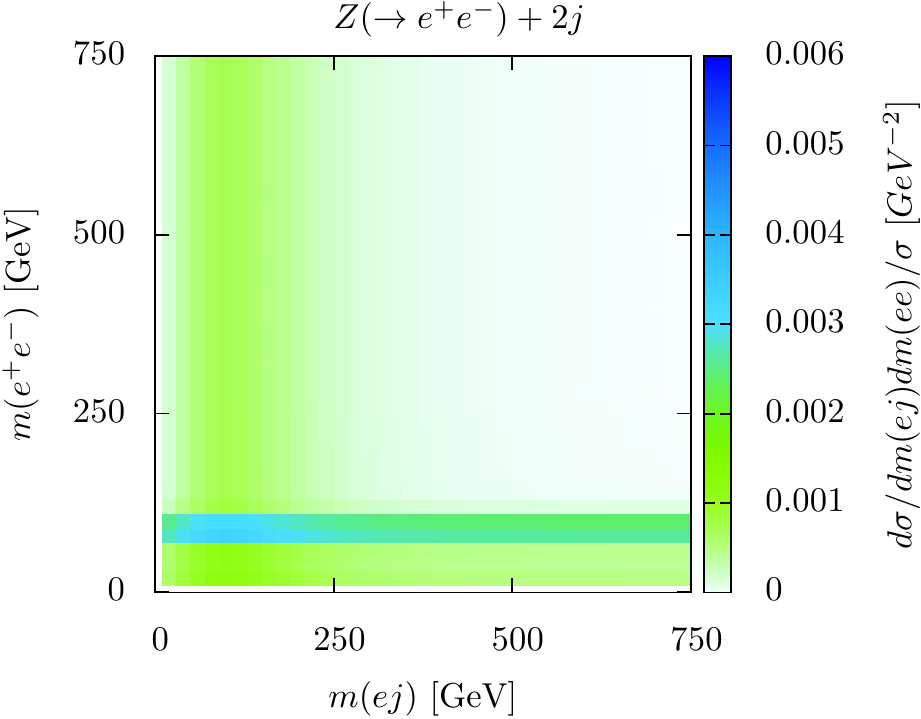} 
\caption{Correlated distributions for charged leptogluon pair production giving rise
to $2j+2l$ final-states (top panels) and the leading 
Standard Model background $pp \to Z +$ jets (bottom panels). We explore the correlations 
between the hardest lepton and jet transverse momenta (left) 
and the invariant 
masses of the dilepton and the hardest jet-lepton systems (right).}
\label{fig:correl-e8}
\end{figure}

\subsection{Signal vs background}
\label{sec:signal}

The signal and backgrounds in this section we simulate with {\sc MadGraph5}~\cite{mg5}. A minimal extension of
the Standard Model including the leptogluon fields and corresponding interactions we  
implement via {\sc FeynRules}~\cite{feynrules}, which also provides the corresponding 
UFO model files~\cite{ufo}. We use {\sc MadAnalysis5 }~\cite{ma5} 
to analyse the event samples.
The effective coupling/scale ratio $a_L/\Lambda$ we fix to $10^{-5}$,
with $a_R=0$. While this choice 
has barely any impact on the total $\mathcal{O}(\alpha_s)$ leptogluon
rates it suppresses 
additional $\mathcal{O}(\alpha_s a^2_{L/R}/\Lambda^2)$ channels
from gluon fusion and a $t$-channel leptogluon exchange.
For the leptogluon mass
the default value is $m_{l_8} = 1$~TeV. For the LHC energy 
we assume $\sqrt{S} = 7$~TeV. 
The leptogluon effective couplings are diagonal in generation space, such that 
we search for electrons and electron-neutrinos,
$l \equiv e^{\pm}, \nu_e$.


\begin{figure}[t!]
\includegraphics[width=0.35\textwidth]{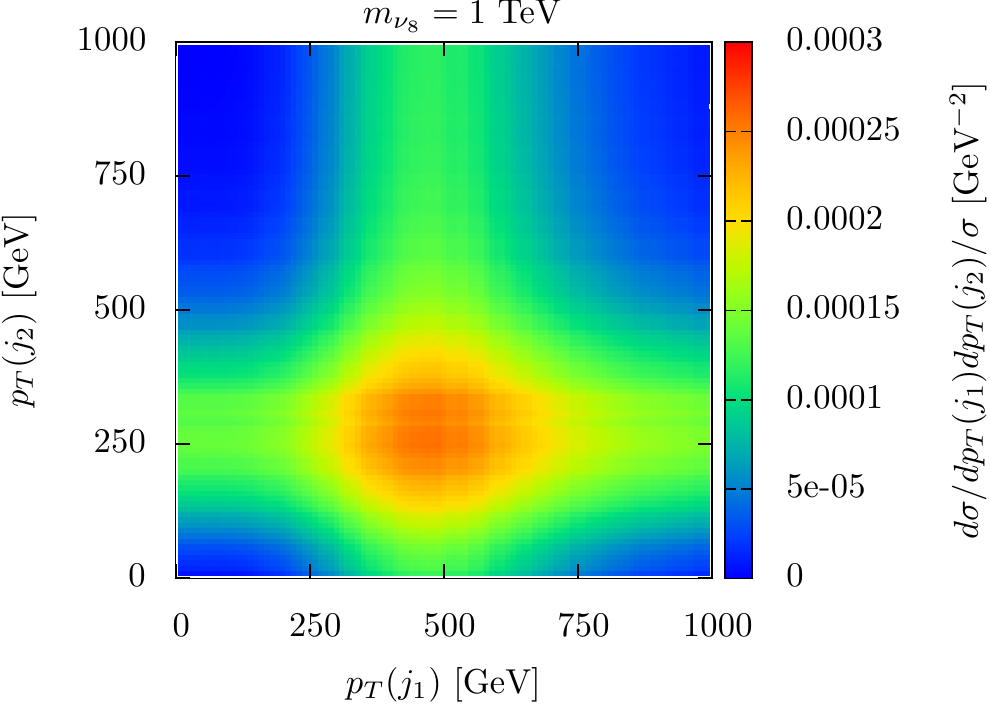}
\hspace*{0.1\textwidth}
\includegraphics[width=0.35\textwidth]{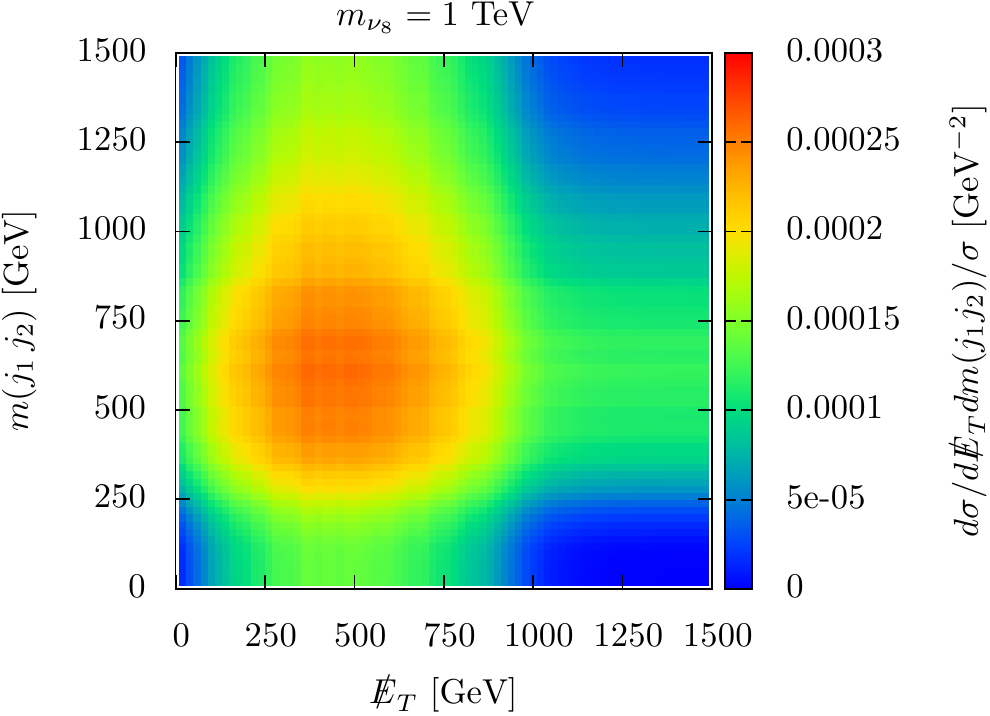} \\ 
\includegraphics[width=0.35\textwidth]{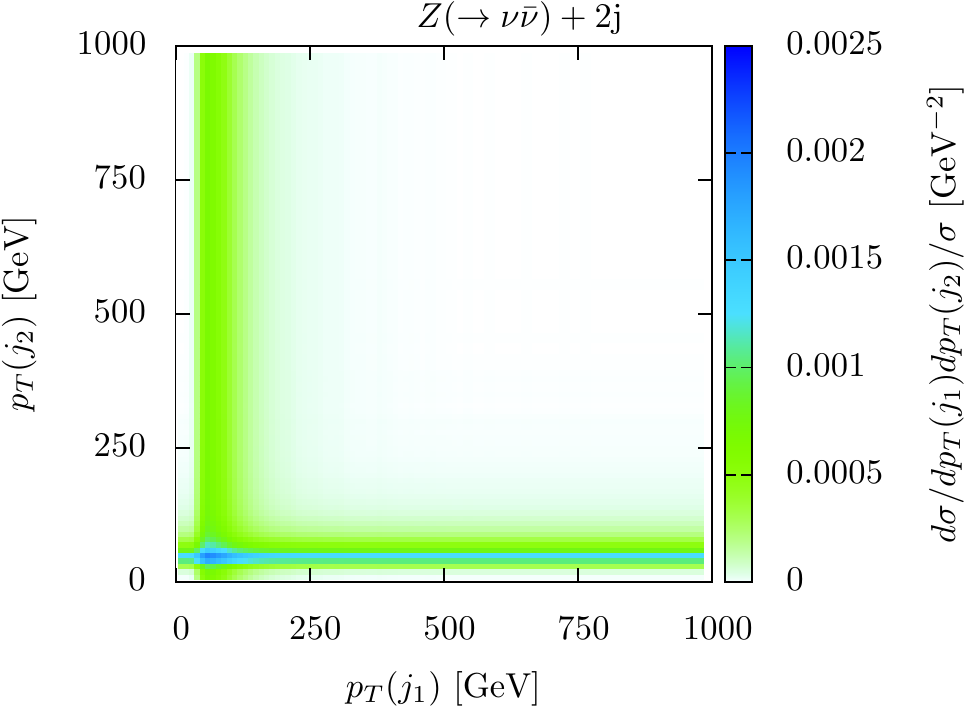}
\hspace*{0.1\textwidth}
\includegraphics[width=0.35\textwidth]{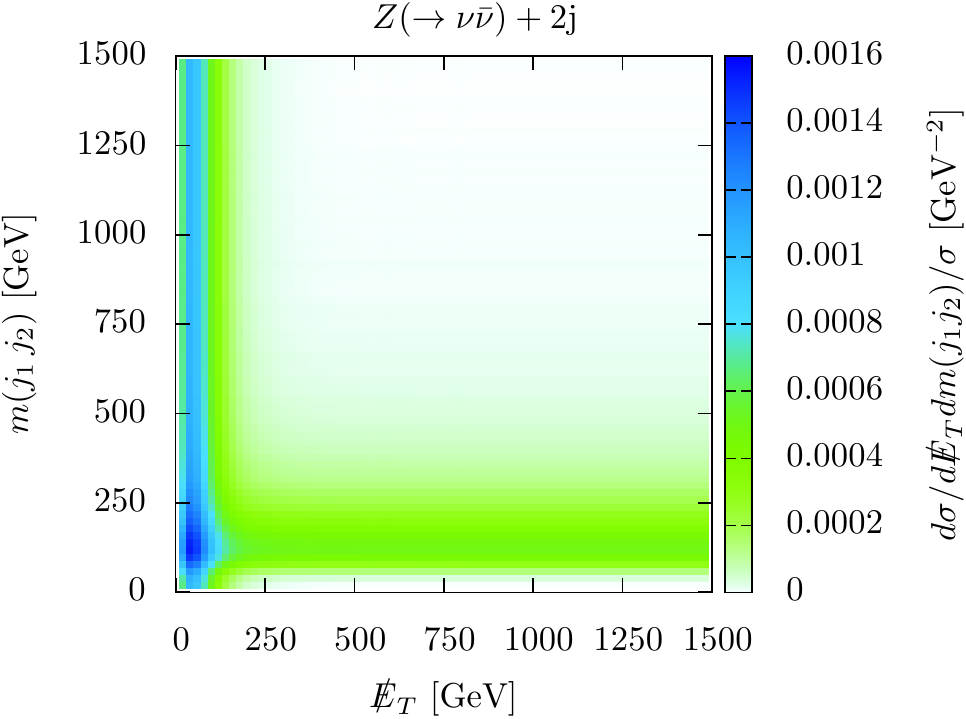} 
\caption{Correlated distributions for neutral leptogluon pair production giving rise
to $2j+\slashed{E}_T$ final-state signatures (top panels) and the leading 
Standard Model background $pp \to Z +$jets (lower panels). We explore the correlations 
between the leading and subleading jet transverse momenta (left)  
and the missing transverse energy vs the invariant 
mass of the leading dijet system (right).}
\label{fig:correl-nu8}
\end{figure}

\smallskip{}
As this process involves the production of colored heavy fields, 
we expect 
a sizable influence
from the initial and final-state QCD radiation. This
gives rise to 
multi-jet configurations which depart from the mere LO expectations 
and leave us with the separation of jets into QCD radiation and decay jets~\cite{sgluon,madgraph_merging,Dreiner:2012sh,autofocus}.
We model QCD jet radiation using MLM jet merging as implemented in
{\sc MadGraph5} to generate 
signal samples including up to two extra hard jets.
Technically, 
we enforce
 $m_{ll}^\text{min} = m_{jj}^\text{min} = 10$~GeV, 
to avoid phase space singularities 
in the backgrounds.

\smallskip{} 
Representative signal and background correlations between different 
distributions involving the hardest lepton and jet decay products
we examine in Figure~\ref{fig:correl-e8}.
They nicely illustrate
the relevant kinematical features which we can rely on for 
a suitable search strategy. 
A
first telltale imprint of the charged leptogluons
involves hard charged leptons,
each of them stemming from the decay of one of the
leptogluons. This translates into very large
dilepton $m(e^+e^-)$ and dijet $m(j_1j_2)$ invariant masses.
The
natural scale for the signal selection is
$m(e^+e^-) \gtrsim m_Z$. 
The leptogluon mass reconstruction, in turn, should proceed through 
the lepton-jet invariant mass
$m(ej)$ which we also 
display 
in Figure~\ref{fig:correl-e8}. 
%
Another hallmark involves 
the transverse momenta
of the hardest jet and
the outgoing charged leptons. 
Additional jet emission from 
the parton shower is responsible for a smearing of this peak.
Since this feature 
is correlated with the 
leptogluon mass it will depart very visibly
from the background structure, as shown in the bottom-left panel
of Figure~\ref{fig:correl-e8}.

\begin{figure}[b!]
\includegraphics[width=0.22\textwidth]{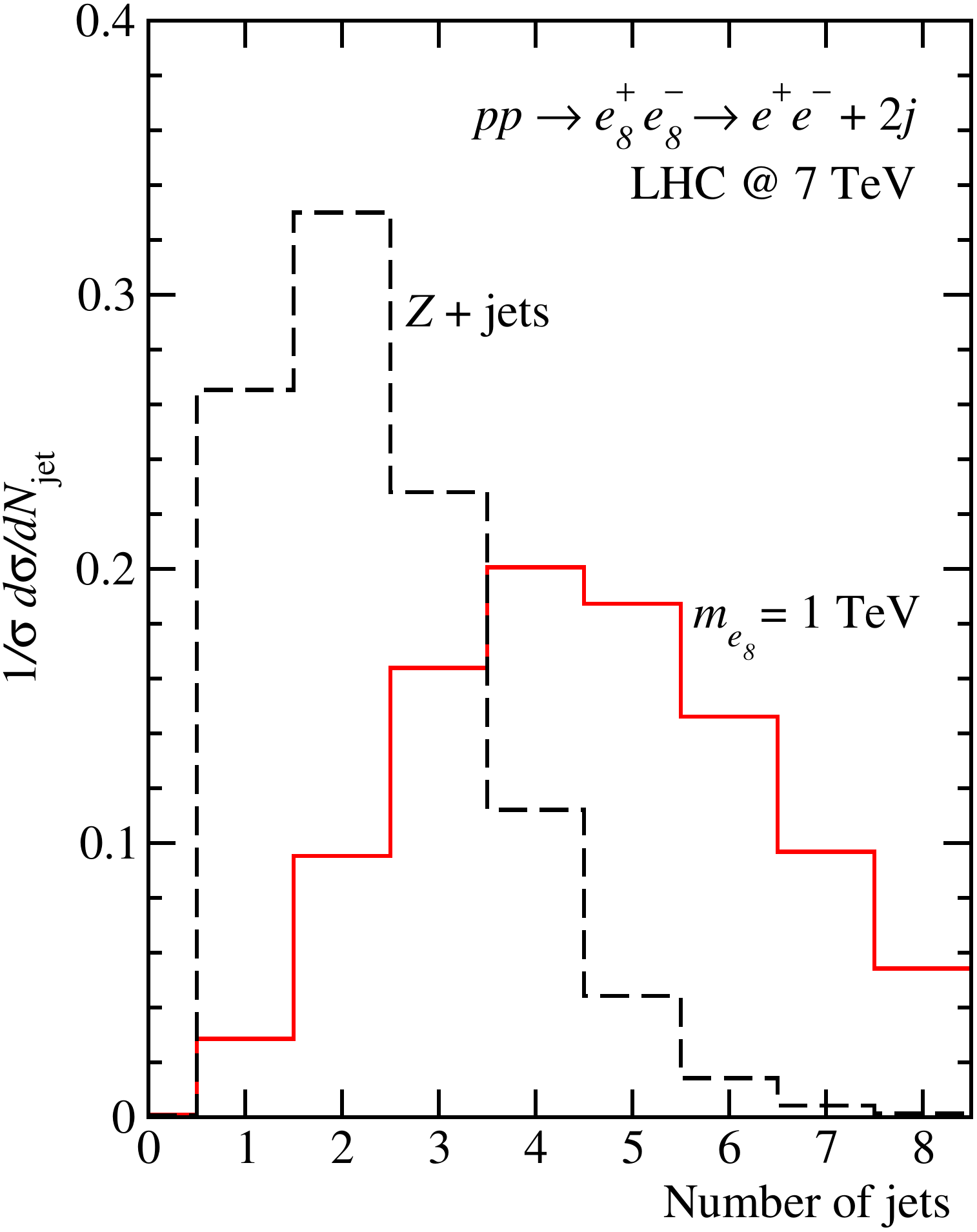} 
\hspace*{0.01\textwidth}
\includegraphics[width=0.235\textwidth]{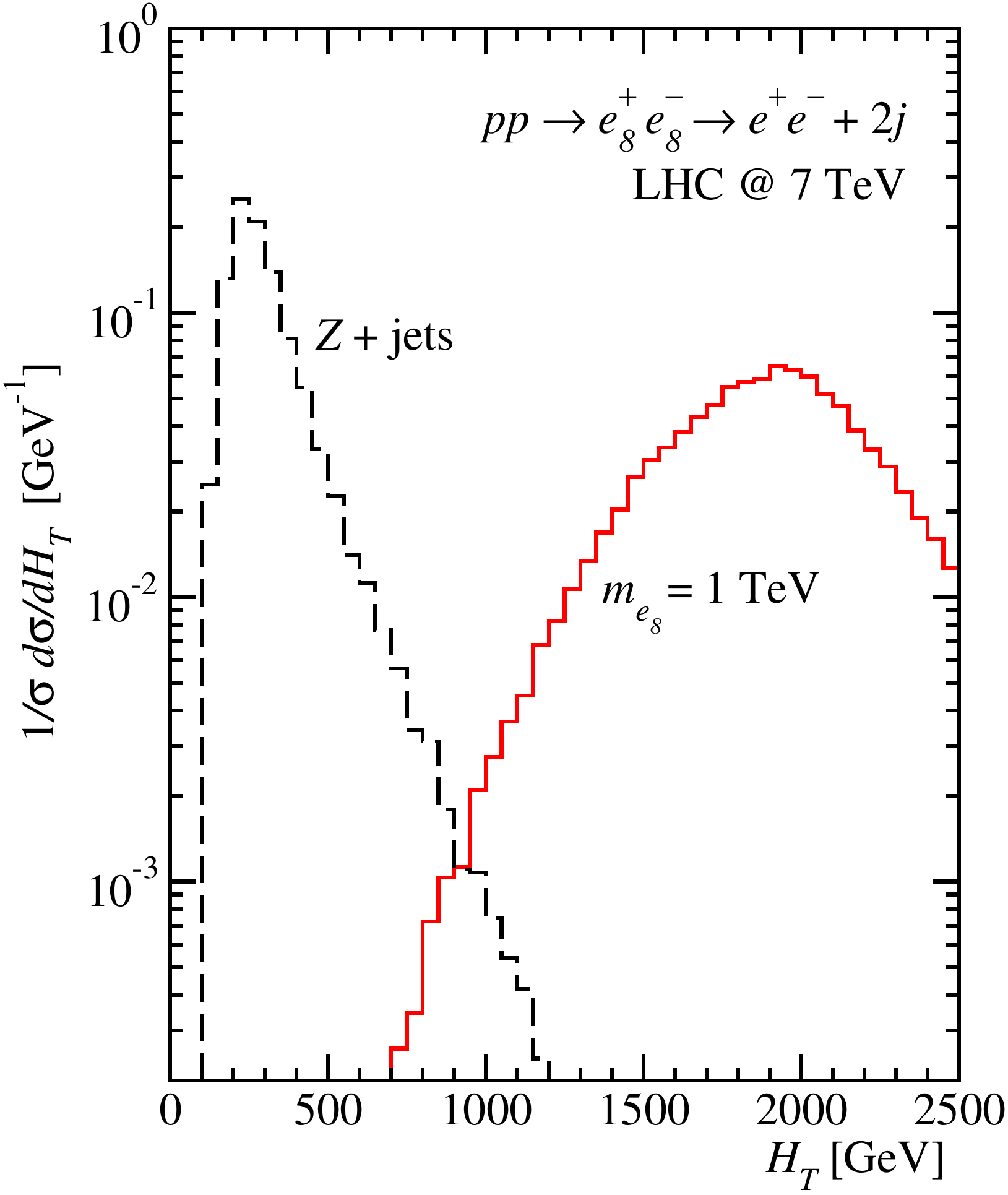} 
\hspace*{0.01\textwidth}
\includegraphics[width=0.22\textwidth]{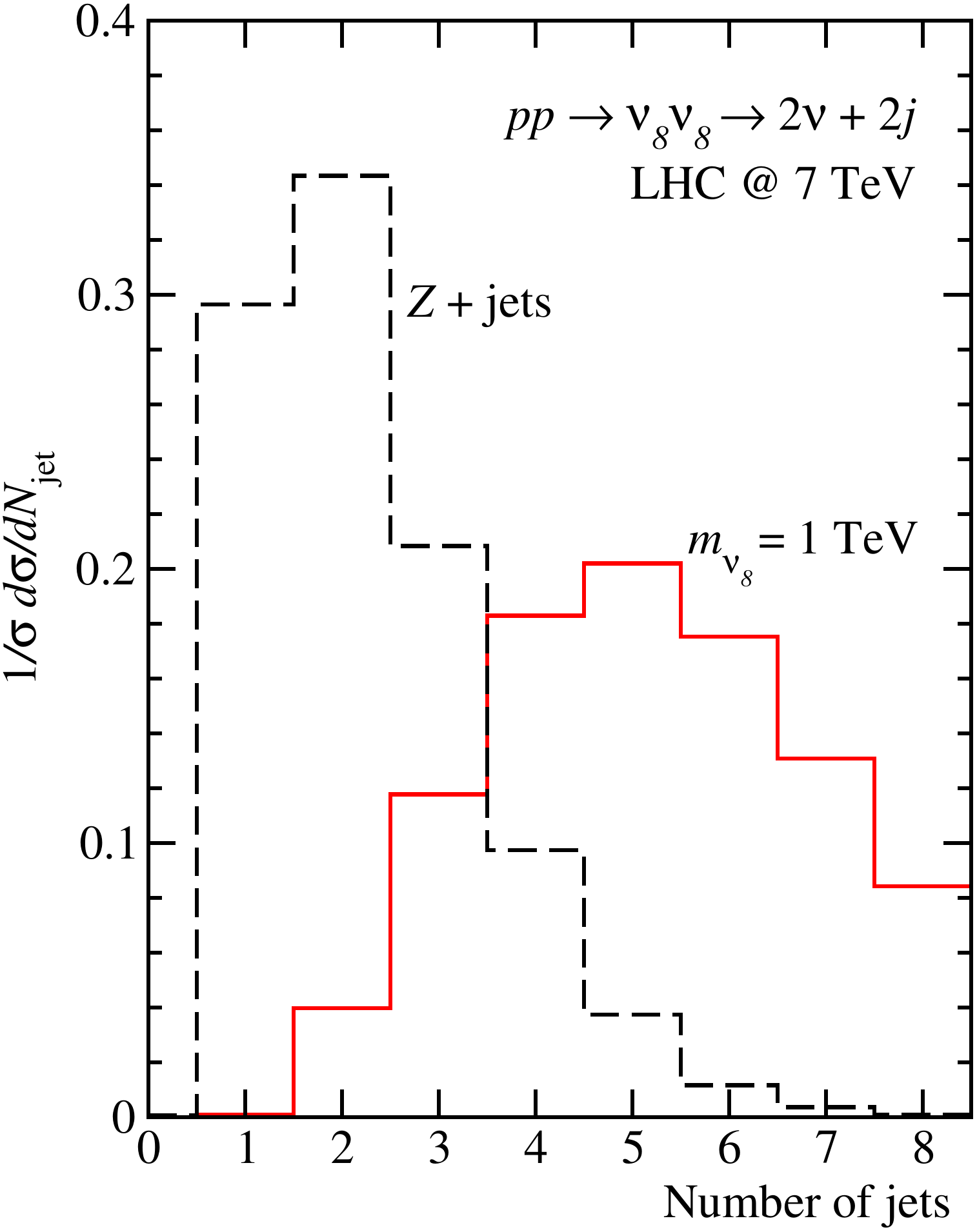} 
\hspace*{0.01\textwidth}
\includegraphics[width=0.235\textwidth]{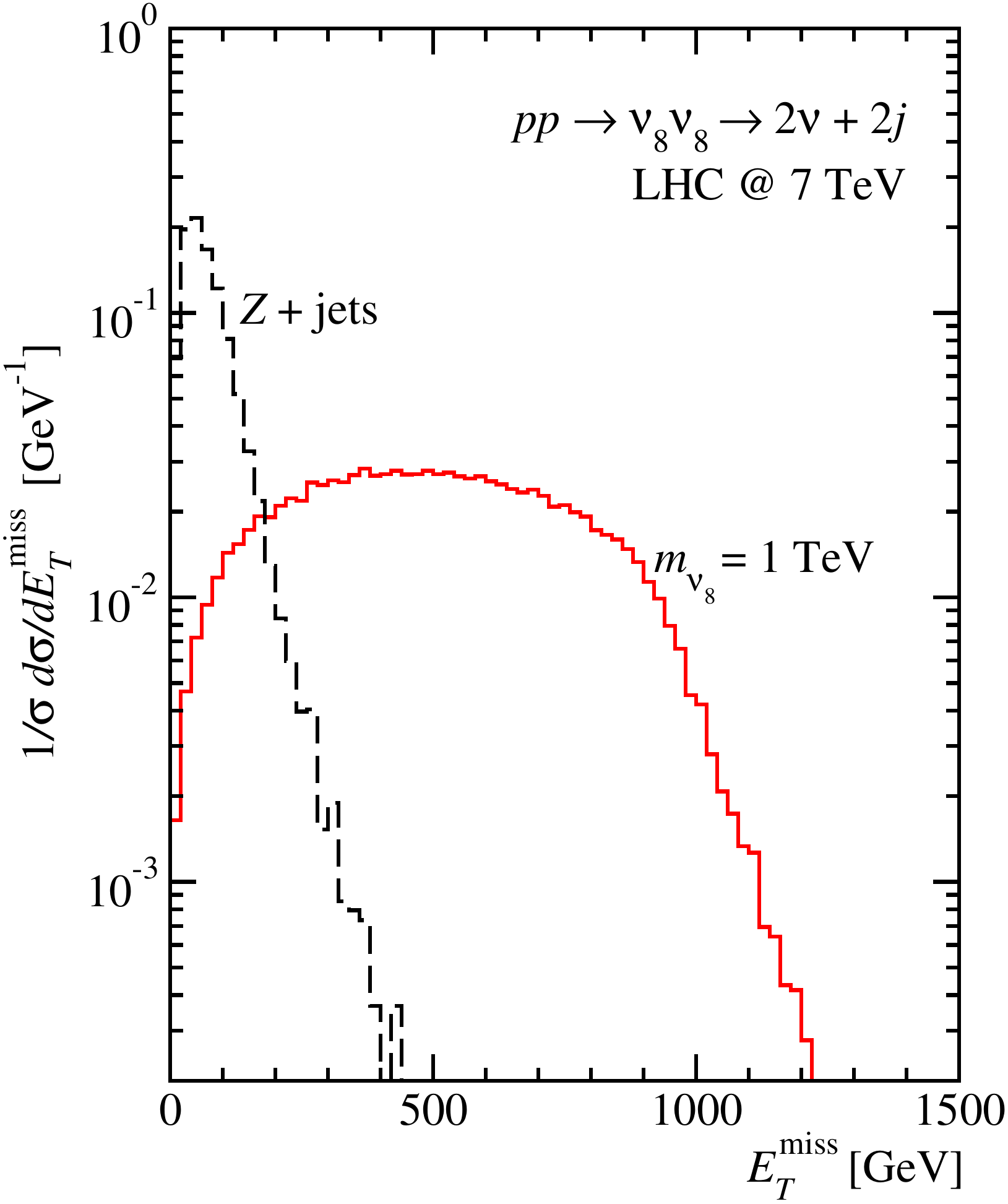} 
\caption{Differential cross sections as a function of
the jet multiplicity 
and the total 
visible (resp. missing) transverse energy for charged (left panels) and neutral (right panels) leptogluons. We also include the 
leading background $pp \to Z +$ jets.
All the histograms we normalize to unity.}
\label{fig:njets-ht}
\end{figure}

\smallskip{}
In analogy to the charged leptogluon case we show the relevant correlated distributions
for the neutral leptogluon signal $pp \to \nu_8\nu_8 \to \slashed{E}_T +$jets 
in Figure~\ref{fig:correl-nu8}. 
In the absence of identified leptons
the experimental prospects are more challenging. 
After using a lepton veto 
to suppress the $t\bar{t}$ and $(Z/\gamma^* \to e^+e^-)+$jets
backgrounds
the leading irreducible background is
$Z +$ jets production with an invisible decay $Z \to \nu\nu$.
The key observable 
is missing transverse energy $\slashed{E}_T$. 
The expected $\slashed{E}_T$ from $Z+$jets production
displays a much softer profile than the signal. 
We illustrate these features 
in Figure~\ref{fig:correl-nu8}.

\smallskip{} 
More generally, signal and backgrounds substantially deviate from each other 
in the number of final-state jets. Owing to the 
$SU(3)_c$ adjoint charge of the leptogluons as well as to their 
large masses
we 
expect the leptogluon events to yield larger jet multiplicities,
as reflected in Figure~\ref{fig:njets-ht}. By the same token, we also retrieve
a much harder total visible energy 
$H_T \equiv \sum_j\,p^j_{T} + p_{T}(e^+)+ p_{T}(e^-)$, whose peak follows once more the 
leptogluon mass.

\begin{figure}[t!]
\includegraphics[width=0.4\textwidth]{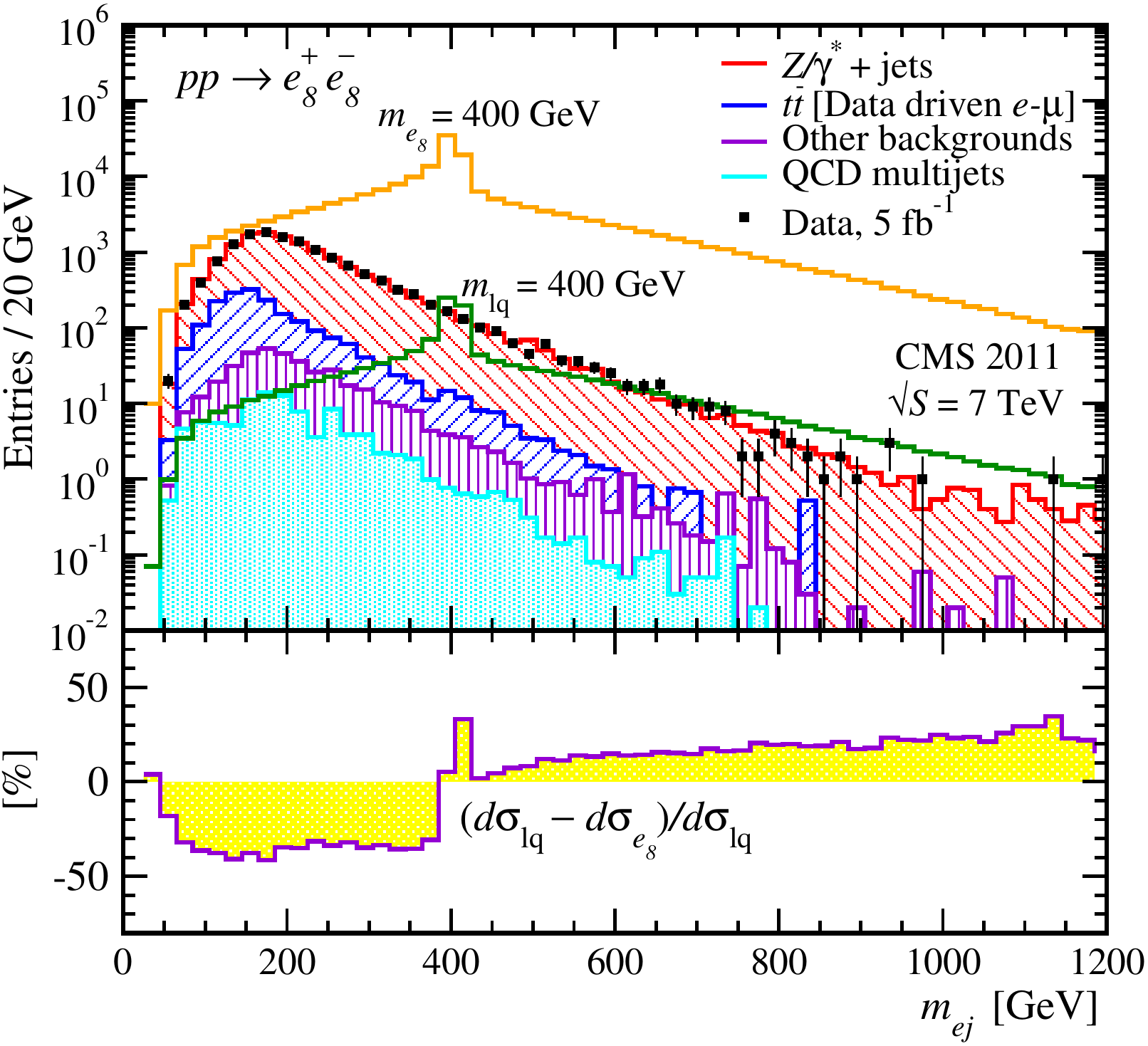}
\hspace{0.05\textwidth}
\includegraphics[width=0.4\textwidth]{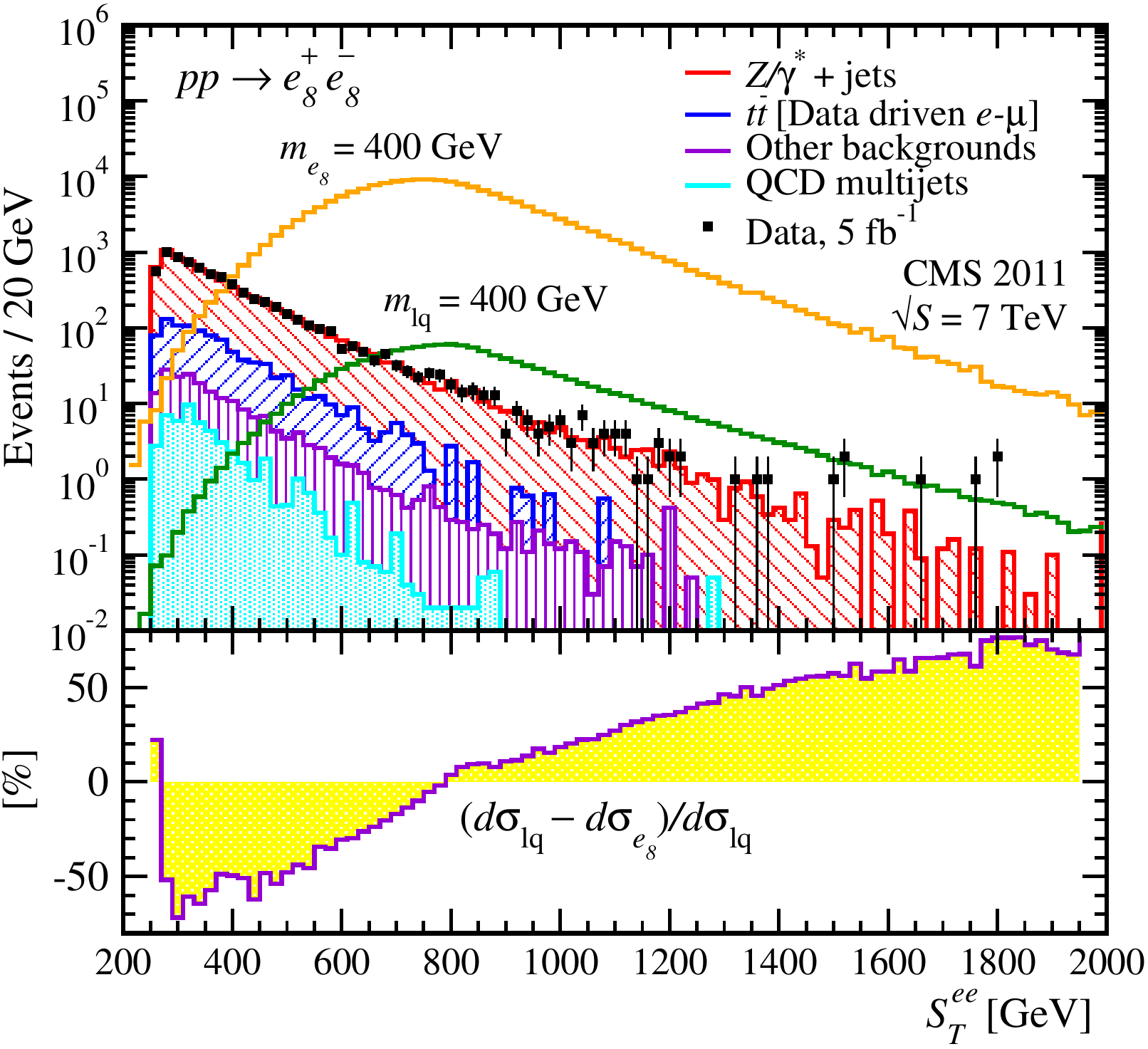} 
\caption{Distributions for the lepton-jet invariant mass
(left) and the scalar transverse energy 
(right) for
leptogluon pair production
followed a decay into a lepton and a jet.
The 
backgrounds we display as cumulative histograms and are taken from ~\cite{:2012dnb,cmspriv}. 
The
leptogluon signal is reweighted to the NLO prediction. 
For comparison, we also include the leptoquark signal 
for the same mass. The bottom panels
illustrate the relative differences between the leptogluon and
leptoquark distributions.}
\label{fig:data1}
\end{figure}

\subsection{Data}
\label{sec:data}

After establishing the NLO description of leptogluon pair production and studying the main kinematic features of the signal and background we are now ready to compare the precision 
predictions to LHC data. 
Specifically, we rely
on the leptoquark searches by CMS~\cite{:2012dnb}.
Leptoquarks (lq) are described by a minimal extension of the Standard Model
including one additional color triplet scalar.
Our discussion henceforth will
cover charged leptogluons only.
The event samples for the signal are originally generated using {\sc Pythia}~6.422 for a range of leptoquark
masses $m_\text{lq} = 250 - 900$~GeV, with a central renormalization
and factorization scale $\mu = m_\text{lq}$. 
The major backgrounds
are determined either from control samples in data or
from Monte Carlo simulations normalized to data in selected control regions. The available experimental
data corresponds to the $\sqrt{S} = 7$~TeV run with 5 fb$^{-1}$ of integrated luminosity.
For further
details we refer the reader to Ref.~\cite{:2012dnb}. 

\bigskip{}
We now replace the leptoquark signal in the CMS leptoquark search with simulated 
leptogluon events,
reweighted to the NLO rate
for each assumed leptogluon mass.
Following Ref.~\cite{:2012dnb},  
we model the triggers
by a set of initial acceptance cuts. 
They are 
tailored to select potential  
$e^+e^-jj$ candidates among those events with two hard  
central leptons and QCD jets. Moreover, the 
jets are required to have a spatial separation of
with respect to the electron candidates, while the 
dilepton invariant mass is enforced to fulfill 
\begin{alignat}{9}
p_T(e) & > 40~\gev \qqquad & |\eta_e| &< 2.5 
                   \qqquad & m(e^+e^-) & > 60~\gev 
                   \qqquad & S_T^{ee} & > 250~\gev \notag \\
p_T(j) & > 30~\gev \qqquad & |\eta_j| &< 2.4 \qqquad & \Delta R_{ej} &> 0.3 \; .
\label{eq:cuts}
\end{alignat}
The cut on the scalar transverse energy $S_T^{ee} =
p_T(e^+)+p_T(e^-)+p_T(j_1)+p_T(j_2)$ significantly reduces the
combinatorial backgrounds.\bigskip

%

%
%

\begin{figure}[t!]
\includegraphics[width=0.5\textwidth]{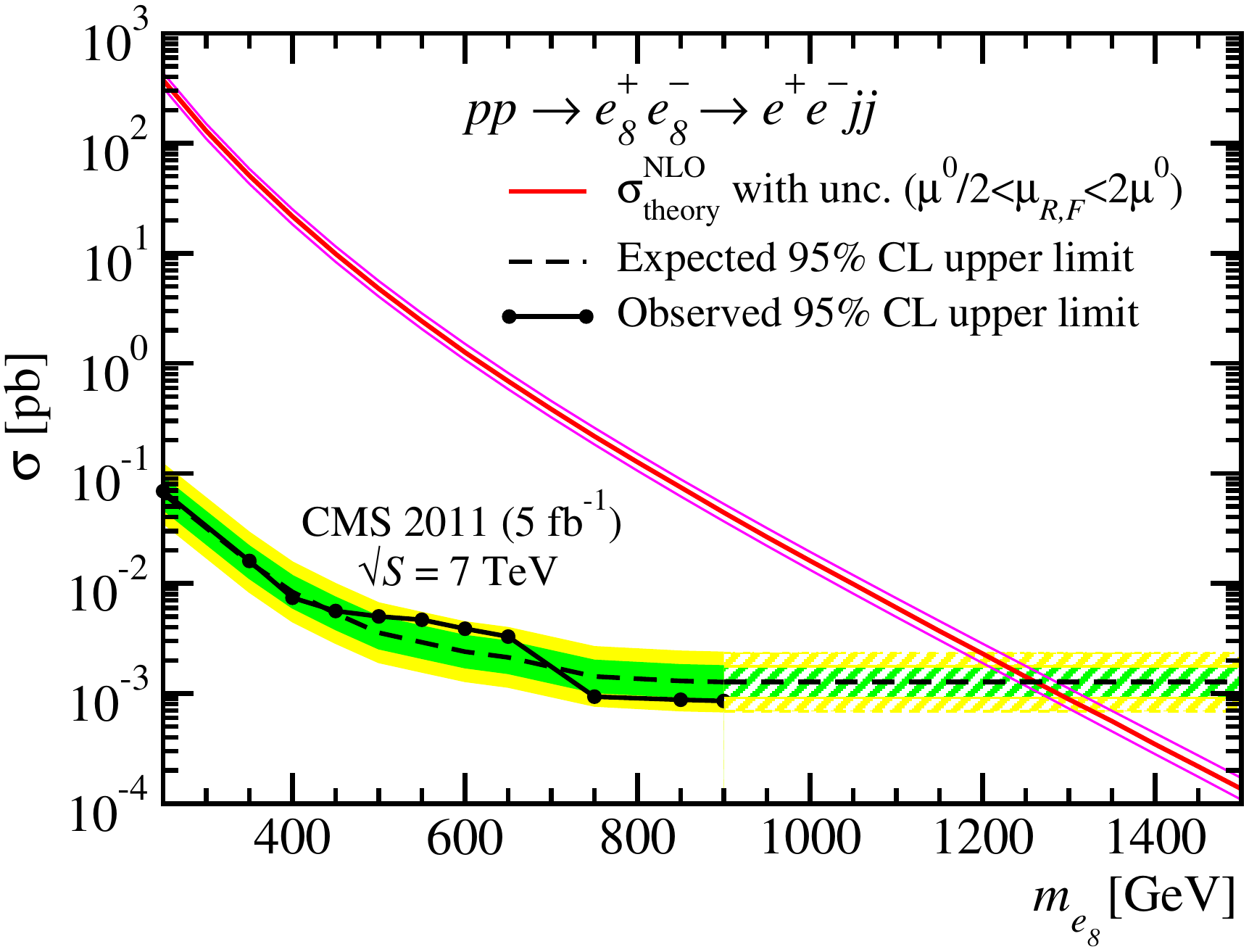} 
\caption{Expected and observed 95\%~CL limits on leptogluon pair production,
followed by a decay into a lepton and a jet, based on the CMS leptoquark searches \cite{:2012dnb,cmspriv}.
The uncertainty bands represent 68\% and 95\%~CL. We overlay the predicted production rates to NLO, including the theoretical uncertainty estimated from a full scale variation.}
\label{fig:data2}
\end{figure}

After this selection the data is well compliant with the Standard Model background expectations,
as reflected in Figures~\ref{fig:data1} and \ref{fig:data2}. 
Figure~\ref{fig:data1} shows 
the electron-jet invariant mass $m_{ej}$ (left)
and the scalar transverse energy $S_T^{ee}$
(right) after acceptance cuts. Data we display as black circles while the backgrounds
form cumulative histograms and are taken from ~\cite{:2012dnb,cmspriv}. Finally, we overlay the
leptogluon and leptoquark signals (solid lines), 
normalized to the respective NLO rates~\cite{Kramer:2004df}.
The masses of both resonances we fix to $m = 400$~GeV, to provide
a replica of Figure~2 from Ref.~\cite{:2012dnb} including the leptogluon pair signal. 
In the lower panels we show the 
relative bin-by-bin differences between the leptoquark and the leptogluon predictions for identical masses.
To deal with combinatorial issues when reconstructing the leptogluon mass,
we distinguish two kinematic regions $\phi \in [0, \pm \pi]$ and we pair
each of the leptons within a given event to the highest $p_T$ jet of
the same hemisphere.




\medskip{}
Adopting the leptoquark analysis we can derive an 
approximate leptogluon mass.
Our starting point is the acceptance--reconstruction efficiency factor 
as a function of the leptoquark mass, provided by CMS~\cite{cmspriv}.
This we combine with the leptoquark acceptances
to get the corresponding mass-dependent efficiencies
$\epsilon_{\text lq}$. The acceptances we estimate
resorting to a leptoquark simulation within {\sc MadGraph5} 
and implementing the 
cuts in Eq.\eqref{eq:cuts}. The same strategy we follow to evaluate
the corresponding leptogluon acceptances.
Finally, we assume that the leptoquark 
efficiencies $\epsilon_\text{lq}$ can be exported to leptogluons,
$\epsilon_\text{lq} \simeq \epsilon_{e_8}$, and derive the combined acceptance--reconstruction efficiency factors
for leptogluons as a function of $m_{e_8}$. 

We observe
that the discrepancies between the leptoquark and leptogluon acceptances,
and so on the resulting
acceptance--efficiency factors, are indeed small. 
In spite of the intrinsically distinct
kinematics of color--triplet scalars vs 
color--octet fermions, non-negligible differences arise primarily in phase space regions hardly
passing the cuts of Eq.\eqref{eq:cuts}. They mostly
attain tails in the transverse momentum and rapidity distributions.
We study these differences for a number
of representative distributions and find them typically at the $\mathcal{O}(10)\%$ level.
This is also indicative that $\epsilon_\text{lq} \simeq \epsilon_{e_8}$
is a reasonable hypothesis.

Finally, we take the CMS leptoquark limits
from a modified frequentist confidence level (CL) analysis~\cite{cls}
and translate them into the leptogluon limit, 
extending them to leptogluon masses $m_{e_8} = 900~\gev - 1.5$~TeV.
This procedure is valid so long as 
the reconstruction efficiency does not
change significantly within this mass range. 
This assumption is justified given the rather general analysis strategy and the experimental setup~\cite{cmspriv}.\bigskip

In Figure~\ref{fig:data2} we show 
the expected and observed $95\%$~CL upper limits on the rate for charged leptogluon pair
production as a function of the mass. 
The uncertainty bands around the median expected limits
correspond to 1 and 2$\sigma$. We overlay a two-sided magenta band
accounting for 
the theoretical uncertainties using the NLO predictions, derived from the independent
variation of the renormalization and
factorization scales around the central value, $\mu^0/2 < \mu < 2\mu^0$.
The intersection of the expected and the predicted rates defines a $2\sigma$ 
exclusion of $m_{e_8} \lesssim 1.2 - 1.3$~TeV. 
This result constitutes a major
improvement of the leptogluon mass bounds quoted at present~\cite{pdg}.

\section{Summary}
\label{sec:sum}

We have studied the pair production of color-octet
leptons at the LHC. 
These leptogluons constitute a minimal extension of
the Standard Model by charged $e^{\pm}_8$ and neutral
$\nu_8$ fields coupling 
to quarks and leptons via higher-dimensional operators.\smallskip

In the first part of the paper we have reported on the complete NLO
calculation of the LHC production rate, using {\sc MadGolem}.  The
genuine new physics structures like Catani--Seymour dipoles and field
as well as mass renormalization are part of the automated setup.  We
find large production rates around $\mathcal{O}(10)$~fb for $\sqrt{S}
= 8$~TeV and leptogluon masses in the TeV range.  Significant quantum
corrections arise mostly from pure QCD, \ie virtual gluons and the
massless initial/final state radiation.  Leptogluon-mediated effects
are mass-suppressed.  
The remaining theoretical uncertainty 
reduces from $\sim 65 \%$ at LO down to $\sim 30 \%$ at NLO.
The NLO predictions are stable across the relevant
phase space regions and agree well with results using LO multi-jet
merging.  This allows us to re-weight event simulations using jet
merging to the NLO total cross section.

In the second part we have studied LHC signatures of leptogluon pair
production. We have described the leptogluon couplings to matter
fields in terms of a dimension 5, generation-diagonal $l_8lg$
chromomagnetic interaction.  Leptogluon pairs appear as characteristic
dijet plus dilepton signatures, featuring hard transverse
momenta and large invariant masses.  These quantities can easily be
distinguished from the irreducible background, mainly governed by $pp
\to Z +$jets.  For charged leptogluons, the signature is the same to
leptoquarks.

Finally, we use this fact to promote the current CMS leptoquark
searches to leptogluon searches. We exclude the existence of charged
leptogluons in the above framework with masses below $\sim 1.2 -
1.3$~TeV.\smallskip

Aside from its phenomenological relevance our work shows how the
automated {\sc MadGolem} framework can be used to efficiently study
new physics models at the LHC.  It completely automatizes the
calculation of NLO cross sections and distributions for pair
production of new particles.  {\sc MadGolem} is an independent modular
add-on to {\sc MadGraph} and can easily be interfaced with its multiple
user options and analysis tools.  Following the present final testing
phase we will make the code available to the theoretical and
experimental communities at the LHC.

\acknowledgments

The work presented here has been in part supported by the Concerted
 Research action 
``Supersymmetric Models and their Signatures at the Large Hadron
 Collider'' and the Strategic Research Program ``High Energy Physics''
of the Vrije Universiteit Brussel (VUB) and 
by the Belgian Federal Science Policy Office through the Interuniversity
Attraction Pole IAP VI/11. DG acknowledges support by the International Max Planck
Research School for Precision Tests of Fundamental Symmetries. It is our pleasure
to deeply thank our CMS experimental colleagues, very specially the CMS
group at VUB, for plenty of enlightening discussions and for providing
us with the experimental data and the background simulation 
displayed in Figures~\ref{fig:data1} and~\ref{fig:data2}.


\section*{Appendix: renormalization}

\paragraph*{Basic setup}
The ultraviolet renormalization counter terms we generate automatically 
from the {\sc Qgraf} leading-order amplitude. The respective field, mass
and $g_s$ renormalization constants 
we conventionally phrase in terms
of two-point functions.
They are part of the leptogluon  model implementation and supplied
as a separate library. 
At present, {\sc MadGolem} fully supports the calculation of NLO QCD corrections for the Standard Model, the MSSM
and several other extensions of the Standard Model featuring heavy colored resonances (e.g. sgluons).
All the necessary renormalization
constants we define
through the relation between the bare 
and the renormalized fields, masses and couplings -- the former 
being denoted with a  ``0'' superscript:
%
%
\begin{equation}
\Psi^{(0)} \to Z^{1/2}_{\Psi}\,\Psi \qqquad
m_{\Psi}^{(0)}  \to m_{\Psi} + \delta\,m_{\Psi} \qqquad
g_s^{(0)} \to g_s + \delta\,g_s
\label{eq:defRC},  
\end{equation}
where $\Psi$ stands for all the strongly interacting degrees of freedom of the 
model, in our case $\Psi = q,g,l_8$.

Given a generic QCD interaction
with Lagrangian density
$\lag(\Psi,m_{\Psi}, g_s)$, we define its associated counterterm 
$\delta\,\lag (\Psi,m_{\Psi}, g_s, \delta\Psi,\delta m_{\Psi}, \delta g_s)$ as
\begin{equation}
\lag(\Psi^{(0)},m^{(0)}_{\Psi}, g^{(0)}_s) = \lag (\Psi,m_{\Psi}, g_s) + \delta\,\lag (\Psi,m_{\Psi}, g_s, \delta\Psi,\delta m_{\Psi}, \delta g_s)
\label{eq:defCT}.
\end{equation}
For further details on our notation setup, we refer the reader to Appendix D of 
Ref.~\cite{susypair}.\bigskip

\paragraph*{Renormalization of the strong coupling constant}

In the presence of the leptogluon field, 
the $\beta$-function of the Standard Model, and thereby the
strong coupling constant renormalization, get modified. 
Using standard notation, we may formally
express the quantum corrections to the quark-gluon
$(q\bar{q}g)$ vertex
in terms of the gluon ($Z_3$) and quark ($Z_2$) field-strength renormalization constants:
\begin{equation}
 Z_1 = Z_g\,Z_2\,Z_3^{1/2}.
\label{eq:z1}
\end{equation}
These factors we can expand to order $\mathcal{O}(\alpha_s)$ as $Z_i = 1 + \delta_i + \mathcal{O}(\alpha_s^2)$,
with the counterterms $\delta_i$ being conventionally defined in the \msbar\, scheme.
The strong coupling constant renormalization at one loop we can thus write as
\begin{equation}
 \delta\,g_s = \delta_1 - \delta_2 - \frac{1}{2}\delta_3
\label{eq:rengs1}.
\end{equation}
Leptogluons only couple to matter fields through higher-dimensional
operators, such as the leptogluon-lepton-gluon interaction described by Eq.\eqref{eq:hdim}.
Consequently, quark self-energies -- and hence $\delta_2$ -- are not affected
by the presence of the leptogluons. 
The fermionic contribution to the $\beta$ function remains therefore unaltered with respect
to that of the Standard Model: 
\begin{equation}
\delta_2^{\msbar} = -\frac{\alpha_s}{4\pi}\,C_F\,\Delta_\epsilon
\label{eq:delta2}.
\end{equation}
The shifted pole in the \msbar\, prescription is $\Delta \equiv \frac{1}{\epsilon}\, - \gamma_E + \log(4\pi)$. 
The same reason explains the absence of leptogluon-mediated corrections
to the quark-gluon vertex at one-loop, 
which means that
\begin{equation}
 \delta_1 = \delta_1^\text{SM} = -\frac{\alpha_s}{4\pi}\,(C_A+C_F)\,\Delta_{\epsilon}
\label{eq:delta1}. 
\end{equation}

\begin{figure}[t!]
\includegraphics[width=0.18\textwidth]{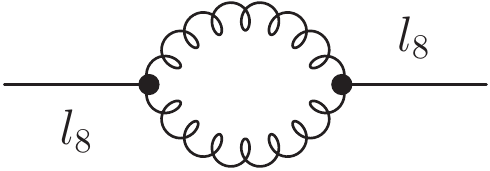} 
\hspace*{0.1\textwidth}
\includegraphics[width=0.16\textwidth]{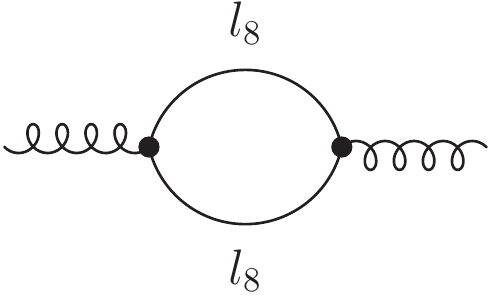} 
\caption{Feynman diagrams for the leptogluon field renormalization at $\mathcal{O}(\alpha_s)$ and for the 
leptogluon-mediated corrections to the gluon field renormalization at $\mathcal{O}(\alpha_s)$.}
\label{fig:2point}
\end{figure}

Instead, the pure QCD gluon/leptogluon interaction
furnishes a novel piece to the gluon self-energies, as displayed in Figure~\ref{fig:2point}.
The latter results into
\begin{equation}
 \delta^{\lo}_3 =  -s\,C_A\,\frac{\alpha_s}{3\pi}\,\Delta_\epsilon
 \label{eq:delta3-leptogluon} 
\end{equation}
with $s=1 (1/2)$ for $l_8 = e^{\pm}_8 (\nu_8)$. 
Adding up the above result to the Standard Model contribution we get
\begin{equation}
 \delta_3^{\msbar} = \delta_3^\text{SM} + \delta_3^{l_8} = \frac{\alpha_s}{4\pi}\,
\left(\frac{5}{3}\,C_A - C_F\,N_F\,T_R\right)
-s\,\frac{\alpha_s}{3\pi}\,C_A\,\Delta_\epsilon
  \label{eq:delta3}.
\end{equation}
The final expression for $\delta\,g_s$
we obtain by plugging Eqs.\eqref{eq:delta1}-\eqref{eq:delta3} 
back into Eq.\eqref{eq:rengs1}, 
while explicitly decoupling the heavy (H)
colored states -- in our case, these are the top-quark
and the leptogluon field(s). We implement such subtraction in 
a conventional zero-momentum
scheme~\cite{prospino_sqgl}, as described in Ref.~\cite{decoupling}.
That way we leave the renormalization
group running of $\alpha_s$ to be merely determined by the light (L)
degrees of freedom. This corresponds to the definition of the measured
value of the strong coupling, for example in a combined fit with the
parton densities. The renormalization constant finally reads
\begin{alignat}{5}
\delta\,g_s = - \frac{\alpha_s}{4\pi}\,\frac{\beta_0^{L} + \beta_0^{H}}{2}\, \Delta_\epsilon 
- \frac{\alpha_s}{4\pi}\,
\left[\frac{1}{3}\,\log\left(\frac{m_t^2}{\mu^2}\right) 
+ 2\,s\,\frac{C_A}{3}\log\left(\frac{m^2_{l_8}}{\mu^2} \right)
\right] \; .
\label{eq:alphas_ct}
\end{alignat}
%
Both
light (L) and heavy (H) colored particles contribute to the coefficient
of the beta function, which in our leptogluon model reads
$\beta_0 = \beta^{L}_0 + \beta_0^{H}$, with
\begin{equation}
\beta_0^L = \left[\frac{11}{3}\,C_A - C_F\,N_F\,T_R \right]
\qquad \text{and} \qquad 
\beta_0^H = -\left[C_F\,T_R + 4\,s\frac{C_A}{3} \right] \; . 
\end{equation}
By default, {\sc MadGolem} sets the number of active flavors to $N_F = 5$.
Standard conventions for the $SU(3)_c$ color factors $C_F = 4/3$,
$C_A = 3$ and $T_{R} = \frac{1}{2}$ are assumed throughout.\bigskip

\paragraph*{Renormalization of the leptogluon sector}

As a massive, color-adjoint particle, the leptogluon 2-point correlation function receives
$\mathcal{O}(\alpha_s)$ corrections due to virtual gluon interchange, as shown
in Figure~\ref{fig:2point}. 
The corresponding UV divergences we can absorb
into a redefinition of the leptogluon mass $m_{l_8}$ and field-strength $\delta\,Z_{l_8}$.
Both quantities we renormalize in the on-shell scheme. The renormalization conditions
require i) the renormalized leptogluon mass to be the pole of the real part of 
the corresponding propagator; and ii) the renormalized propagator to have unit residue.
These conditions we can cast in the following guise:
\begin{eqnarray}
 \Re\text{e}\,\hat{\Sigma}_{l_8}(p^2 = m^2_{l_8}) = 0;
\qquad \,\lim_{p^2 \to m_{l_8}^2}\, \left[\cfrac{\slashed{p}+m_{l_8}}{p^2-m^2_{l_8}}\right] \,
\,\Re{\rm e}\hat{\Sigma}_{l_8}(p)\,\psi_{l_8}(p) = 0.
 \label{eq:OSconditions}
\end{eqnarray}
%
On the other hand, the (real part of the) renormalized leptogluon self-energy, $\Re\text{e}\hat{\Sigma}_{l_8}$,
we conventionally write as:
\begin{equation}
\Re\text{e} \hat{\Sigma}_{\lo}(\slashed{p}) = \Re\text{e}\left[\slashed{p}\,P_L\,\hat{\Sigma}_{\lo, L}(\slashed{p})+
\slashed{p}\,P_R\,\hat{\Sigma}_{\lo, R}(\slashed{p})
+m_{\lo}\Sigma_{l_8, S}(\slashed{p}) \right]
\label{eq:selfleptogluon},
\end{equation}
where the scalar components $L,R$ and $S$ render:
\begin{alignat}{5}
\hat{\Sigma}_{\lo, L/R}(\slashed{p}) & =  \Sigma_{\lo, L/R}(\slashed{p}) + \delta\,Z_{\lo, L/R} \\
\hat{\Sigma}_{\lo, S}(\slashed{p}) & = \Sigma_{\lo, S}(\slashed{p})-\frac{1}{2}\,(\delta\,Z_{\lo,L}+\delta\,Z_{\lo,R}) - \frac{\delta\,m_{\lo}}{m_{\lo}}
\label{eq:selfcomponents}.
\end{alignat}
Enforcing the on-shell renormalization conditions Eq.\eqref{eq:OSconditions} on the above
leptogluon 2-point functions, we can finally anchor the field and mass renormalization constants
as follows:
\begin{eqnarray}
\delta Z_{l_8,L/R} &=& -\Re\text{e}\,\rself_{l_8, \,L/R}\,(m^2_{l_8}) - m^2_{l_8}\,\Re\text{e}\,
\left[\Sigma{'}_{l_8, L}(m_{l_8}^2) + \Sigma{'}_{l_8, R}(m_{l_8}^2) + 2\,\Sigma{'}_{l_8, S}(m_{l_8}^2)  \right]
\nonumber \\
\frac{\delta\,m_{l_8}}{m_{l_8}} &=& \frac{1}{2} \Re\text{e}\, 
\left[\Sigma_{l_8, L}(m_{l_8}^2) + \Sigma_{l_8, R}(m_{l_8}^2) + 2\Sigma_{l_8, S}(m_{l_8}^2) \right] 
 \label{eq:leptogluonOS}.
\end{eqnarray}
Notice that the above expressions Eq.\eqref{eq:selfleptogluon}-Eq.\eqref{eq:leptogluonOS}
equally hold for both charged and neutral leptogluon states. Moreover, as far as we disregard electroweak
effects, in practice we will find no distinction between the respective left and right-handed components.
The conventional shorthand notation $f{'}(p^2) \equiv df(p^2)/dp^2$ is employed therewith.

\smallskip{} 
The analytic form of all renormalization constants
we reduce down to 1 and 2-point scalar one-loop functions,
in the conventions of ~\cite{oneloop}. 
UV divergences
we handle by means of a standard 'tHooft-Veltman
dimensional regularization scheme within $n=4-2\epsilon$ dimensions.
For the leptogluon mass and field strength renormalization, these quantities read:
\begin{eqnarray}
 \delta\,Z_{l_8} &= &-\frac{\alpha_s}{4\pi}\, C_A\,\left[\Re\text{e}\,B_0(m_{l_8}^2,m_{l_8}^2,0)-4\,m_{l_8}^2\,\Re\text{e}\,B'_0(m_{l_8}^2, m_{l_8}^2,0) -2 \right] \label{eq:leptogluonfield} \\
\delta\,m_{l_8} &=& \frac{\alpha_s}{4\pi}\,C_A\,\left[2 - 3\,\Re\text{e}\,B_0(m^2_{l_8},m^2_{l_8},0) \right]. \label{eq:leptogluonmass}
\end{eqnarray}
Unsurprisingly, the above expressions are identical to the gluon-mediated
correction to 
the gluino field and mass renormalization constants. For the gluino case, however, we rely as well
on the additional SUSY-QCD contributions triggered
by the squark interchange.

\smallskip{}
Finally, 
the explicit analytical expressions for the 
different UV counter terms involving the leptogluon interactions $\delta\,\lag$ we quote in Table~\ref{tab:cts} 
in terms of the 
field, mass and strong coupling renormalization constants.

\begin{table}[t!]
\begin{tabular}{ll}  
\myrbox{\includegraphics[scale=0.001]{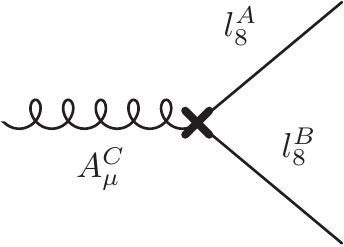}} \phantom{hallooo}
&  $-g_s\,f^{ABC}\,\overline{\psi}^A_{l_8}\, \gamma^\mu\,
   \left[\delta\,g_s + \frac{1}{2}\,\left( \delta\,Z_{l_8} + \delta\,Z_{\bar{l}_8} 
                                         + \delta\,Z_g\right)
   \right] \, \psi^B_{l_8}\,A^C_{\mu}$  \cr
\myrbox{\includegraphics[scale=0.001]{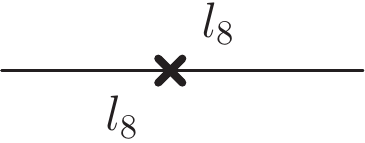}} 
& $i\overline{\psi}_{l_8} (\slashed{p}-m_{l_8})\,\delta\,Z_{l_8}\psi_{l_8}
     - \delta\,m_{l_8}\,\overline{\psi}_{l_8}\psi_{l_8}$
\end{tabular}
\caption{Counter term Feynman rules for the leptogluon-mediated interactions. \label{tab:cts}}
\end{table}

\end{document}